\documentclass{elsarticle}

\usepackage{hyperref}

\journal{Journal of \LaTeX\ Templates}

\usepackage{amssymb}
\usepackage{graphicx}
\usepackage{url}
\usepackage{amsmath}
\usepackage{amssymb}
\usepackage{color}
\usepackage{epsf}
\usepackage{color}
\usepackage{graphicx}
\usepackage{fancybox}
\usepackage{multicol}
\usepackage{hhline}
\usepackage{latexsym}
\usepackage[all]{xy}
\DeclareMathAlphabet{\mathpzc}{OT1}{pzc}{m}{it}
\usepackage{enumerate}

\mathchardef\myhyphen="2D

\newcounter{example-counter}
\newenvironment{example}%
{\vskip \abovedisplayskip \refstepcounter{example-counter}%
\noindent {\bf Example \arabic{example-counter}.}}%

\newtheorem{proposition}{Proposition}
\newtheorem{definition}{Definition}
\newtheorem{lemma}{Lemma}
\newtheorem{theorem}{Theorem}
\newtheorem{corollary}{Corollary}

\newcommand{\boxtheorem}{\hfill $\blacksquare$\vspace{1mm}}
\newcommand{\ignore}[1]{}
\newcommand{\nit}[1]{{\it #1}}

\DeclareMathAlphabet{\mathpzc}{OT1}{pzc}{m}{it}

\newcommand{\eat}[1]{}
\newcommand{\mc}[1]{\mathcal{ #1}}

\ignore{+++
{\vskip \abovedisplayskip \refstepcounter{lemmaA-counter}%
\noindent {\bf Lemma A.\arabic{lemmaA-counter}.}}%

{\vskip \abovedisplayskip \refstepcounter{definitionA-counter}%
\noindent {\bf Definition A.\arabic{definitionA-counter}.}}%

+++}

\newcommand{\defproof}[2]{{\noindent\bf Proof of #1:\
}#2 \boxtheorem\\ \vspace{2mm}}

\newcommand{\hproof}[1]{{\noindent\bf Proof:\ }#1 \boxtheorem}

\newcommand{\red}{}

\newcommand{\bl}[1]{\textcolor{blue}{#1}}
\newcommand{\bblue}[1]{\textcolor{blue}{#1}}
\newcommand{\re}{}

\newcommand{\comlb}[1]{{\vspace{2mm}\noindent \bf \re{\underline{COMM(LEO):}}}~ #1 \hfill {\bf END.}\\}
\newcommand{\mn}{{\!\!-N}}

\newcommand{\dproof}[1]{{\noindent\bf Proof:\
}#1 \boxtheorem}

\ignore{+++
\newcounter{theorem-counter}
\newcounter{corollary-counter}
\newcounter{lemma-counter}
\newcounter{definition-counter}
\newcounter{example-counter}
\newcounter{proposition-counter}
\newcounter{remark-counter}
\newcounter{definitionA-counter}
\newcounter{lemmaA-counter}
\newcounter{propositionA-counter}
\setcounter{lemmaA-counter}{0}

\setcounter{propositionA-counter}{0}
\setcounter{definitionA-counter}{0}

\setcounter{theorem-counter}{0}
\setcounter{corollary-counter}{0}
\setcounter{lemma-counter}{0}
\setcounter{definition-counter}{0}
\setcounter{example-counter}{0}
\setcounter{proposition-counter}{0}
\setcounter{remark-counter}{0}

\newenvironment{theorem}%
{\vskip \abovedisplayskip \refstepcounter{theorem-counter}%
\noindent {\bf Theorem \arabic{theorem-counter}.}}%

\newenvironment{corollary}%
{\vskip \abovedisplayskip \refstepcounter{corollary-counter}%
\noindent {\bf Corollary \arabic{corollary-counter}.}}%

\newenvironment{lemma}%
{\vskip \abovedisplayskip \refstepcounter{lemma-counter}%
\noindent {\bf Lemma \arabic{lemma-counter}.}}%

\newenvironment{definition}%
{\vskip \abovedisplayskip \refstepcounter{definition-counter}%
\noindent {\bf Definition \arabic{definition-counter}.}}%

\newenvironment{example}%
{\vskip \abovedisplayskip \refstepcounter{example-counter}%
\noindent {\bf Example \arabic{example-counter}.}}%

\newenvironment{proposition}%
{\vskip \abovedisplayskip \refstepcounter{proposition-counter}%
\noindent {\bf Proposition \arabic{proposition-counter}.}}%

{\vskip \abovedisplayskip \refstepcounter{remark-counter}%
\noindent {\bf Remark \arabic{remark-counter}.}}%

+++}

\begin{document}

\begin{frontmatter}

\title{{\bf Causes for Query Answers from Databases: Datalog Abduction, View-Updates, and Integrity Constraints}}

\author{{\bf Leopoldo Bertossi}\footnote{ \ {\bf Contact author.} \ Carleton University, \ School of Computer Science, \
Ottawa, Canada. \ Email: \ bertossi@scs.carleton.ca.} \ and \ {\bf Babak Salimi}\footnote{ \ University of Washington, Computer Science \& Engineering, Seattle, USA. Email: \ bsalimi@cs.washington.edu.}
}

\begin{abstract} 
Causality has been recently introduced in databases, to model, characterize, and possibly compute causes
for query answers. Connections between QA-causality  and {\em consistency-based diagnosis} and {\em database repairs}
(wrt. integrity constraint violations) have already been established. In this work we establish precise  connections between
QA-causality   and both {\em abductive diagnosis} and the {\em view-update problem} in databases, allowing us to obtain new algorithmic and complexity
results for QA-causality.  We also obtain new results on the complexity of {\em view-conditioned causality}, and investigate the
notion of QA-causality in the presence of integrity constraints, obtaining complexity results from a connection with view-conditioned causality. The abduction connection under integrity constraints
allows us to obtain algorithmic tools for QA-causality.
\end{abstract}




\end{frontmatter}

\section{Introduction}

Causality is an important concept that appears at the foundations of many scientific disciplines, in the practice of technology,
and also in our everyday life. Causality
is fundamental to understand and manage {\em uncertainty} in data, information, knowledge, and theories.
In data management in particular, there is a need to represent, characterize
 and compute \ignore{the} causes that explain why certain query results are obtained or not, or why natural semantic conditions, such
 as integrity constraints, are satisfied or not. Causality can also
 be used to explain the contents of a view, i.e. of a predicate with virtual
 contents that is defined in terms of other physical, materialized relations (tables).

Most of the work on causality has been developed in the context of artificial intelligence \cite{pearl} and
Statistics \cite{pearlEtal},
 and little has been said about causality in data management.
In this
 work we concentrate on causality as defined for- and applied to relational databases. In a world of big, uncertain data, the necessity to understand the data beyond direct query answers, introducing explanations in different  forms, becomes particularly relevant.

The notion of causality-based explanation for a query result   was introduced in \cite{Meliou2010a}, on the basis of the deeper concept of {\em actual causation}.\footnote{ \ In contrast with general causal claims, such as ``smoking causes cancer", which refer some sort of related events, actual causation  specifies a particular instantiation of a causal relationship, e.g., ``Joe's smoking is a  cause for his cancer".} We will refer to this notion
as {\em query-answer causality} (or simply, QA-causality).
Intuitively, a database atom (or tuple) $\tau$ \ is an {\em actual cause} for an answer $\bar{a}$ to a
conjunctive query $\mc{Q}$ from  a relational  instance $D$ if there is a ``contingent" subset of tuples $\Gamma$, accompanying $\tau$,
such that, after removing $\Gamma$ from $D$, removing $\tau$ from $D\smallsetminus \Gamma$ causes $\bar{a}$ to switch from being an answer to being a non-answer (i.e.
not being an answer). Usually, actual causes and contingent tuples  are restricted to be among a pre-specified set
of {\em endogenous tuples}, which are admissible, possible candidates for causes, as opposed to {\em exogenous tuples}.

A cause $\tau$ may have different associated contingency sets $\Gamma$. Intuitively, the smaller they are the strongest is $\tau$ as a cause (it need less company to
undermine the query answer). So, some causes may be stronger than others. This idea is formally captured through the notion of {\em causal responsibility},
and introduced in \cite{Meliou2010a}. It reflects the relative degree of actual causality. In applications involving large
data sets, it is crucial to rank potential causes according to their responsibilities \cite{Meliou2010b, Meliou2010a}.

Furthermore, {\em view-conditioned causality} (in short, vc-causality) was proposed in \cite{Meliou2010b, Meliou2011} as a restricted form of QA-causality, to determine causes for  unexpected
query results, but  conditioned to the correctness of prior knowledge  that cannot be altered by hypothetical tuple deletions.

Actual causation, as used in  \cite{Meliou2010a, Meliou2010b, Meliou2011}, can be traced back to
\cite{Halpern05}, which provides  a model-based account of causation on the basis of {\em counterfactual dependence}.\footnote{ \ As discussed in \cite{icdt15}, some objections
to the Halpern-Pearl model of causality and the corresponding changes \cite{halpern14,halpern15} do not affect
 results in the context of databases.} {\em Causal responsibility} was introduced  in \cite{Chockler04}, to provide a graded, quantitative notion of causality when multiple causes may over-determine an outcome.

\ignore{Apart from the explicit use of causality, research on explanations for query results has focused mainly, and rather implicitly, on provenance
\cite{BunemanKT02, Cheney09, tannen}, and
more recently, on provenance for non-answers \cite{ChapmanJ09, HuangCDN08}.
}
\ignore{For connections between
causality and provenance, see \cite{Meliou2010a, Meliou2010b}. However, causality is a more refined notion that identifies causes
for query results on the basis of  user-defined criteria, and ranks causes according to
their responsibility \cite{Meliou2010b}.
}

In \cite{icdt15,tocs15} connections were established between QA-causality and {\em database repairs} \cite{2011Bertossi}, which allowed to obtain several complexity results for QA-causality related problems.
Connections between QA-causality and {\em consistency-based diagnosis} \cite{Reiter87} were established
in  \cite{icdt15,tocs15}. More specifically, QA-causality and causal responsibility
were characterized in terms of consistency-based diagnosis, which led to
new algorithmic results for QA-causality \cite{icdt15,tocs15}. In \cite{buda14}  first connections between QA-causality, view updates, and {\em abductive diagnosis} in Datalog \cite{console91, EiterGL95} were announced.
We elaborate  on this in the rest of this section.

The definition of QA-causality applies to monotone queries \cite{Meliou2010a, Meliou2010b}.\footnote{\red{That a query is monotone means that the set of answers may only grow when new tuples are inserted into the database.}} However, all complexity and algorithmic results in \cite{Meliou2010a, icdt15} have
been  restricted to first-order (FO) monotone queries, mainly conjunctive queries. However, Datalog queries \cite{ceri90, Abiteboul95}, which are also monotone, but may contain recursion, require investigation
in the context of QA-causality.

In contrast  to consistency-based diagnoses, which is  usually practiced with  FO specifications, abductive diagnosis is commonly
done with different  sorts of  logic programming-based specifications \cite{DeneckerK02, EiterGL97, Gottlob10b}. In particular, Datalog
can be used as the specification language, giving rise to Datalog-abduction  \cite{Gottlob10b}. In this work we establish a  relationship
between Datalog-abduction and QA-causality, which allows us to obtain complexity results for QA-causality for Datalog queries.

  We also explore fruitful connections between QA-causality and the classical and important {\em view-update problem} in databases \cite{Abiteboul95}, which is
   about updating a database through views. An important aspect of the problem is that one wants the base relations (sometimes called ``the source database") to change in a minimal way
   while still producing the intended view updates. This is
    an update propagation problem, from
views to base relations.

The {\em delete-propagation} problem  \cite{BunemanKT02, Kimelfeld12a, Kimelfeld12b} is a particular case of the view-update problem, where only tuple deletions are allowed
from the views. If the views are defined by monotone queries, only source deletions can give an account of view deletions. When only a subset-minimal set of deletions
from the base relations is expected to be performed, we are in the ``minimal source-side-effect" case. The ``minimum source-side-effect" case appears when that set is required to have a  minimum
cardinality. In a different case, we may want to minimize the side-effects on the
 view, requiring that other tuples in the (virtual) view contents are not affected (deleted)  \cite{BunemanKT02}.

 In this work we provide precise connections between QA-causality and different variants of the delete-propagation problem.
 In particular, we show that the minimal-source-side-effect deletion-problem and the minimum-source-side-effect deletion-problem are related to QA-causality for monotone queries and the {\em most-responsible cause problem}, \red{as
 investigated in \cite{Meliou2010a,icdt15,tocs15}}.  The minimum-view-side-effect deletion-problem
 is related to vc-causality. We establish precise mutual characterizations (reductions) between these problems, obtaining in particular, new complexity results for view-conditioned causality.

 \re{Finally, we also  define and investigate the notion of query-answer causality in the presence of {\em integrity constraints}, which are logical dependencies between database
tuples \cite{Abiteboul95}. Under the assumption that the instance at hand satisfies a given set of ICs, the latter
should have an effect on the causes for a query answer, and their computation. We show
that they do, proposing a notion of QA-cause under ICs. But taking advantage of the connection with Datalog-abduction (this time under ICs on the extensional relations), we develop techniques
to compute causes for query answers from Datalog queries in the presence of ICs.}

Summarizing, our main results  are the following:\ignore{\footnote{ \ Possible connections between QA-causality and delete-propagation  were suggested
 in \cite{buda14}.}} 

 \begin{enumerate}
 \item
  We establish precise connections between QA-causality for Datalog queries  and abductive diagnosis from Datalog specifications, i.e.
  mutual characterizations  and computational reductions between them.

\item
  We establich that, in contrast to (unions of) conjunctive queries, deciding tuple causality
for Datalog queries is \nit{ {NP}}-complete in data.

 \item We identify a class of (possibly recursive) Datalog queries for which deciding causality is \re{fixed-parameter tractable in combined complexity}.

 \item We establish that deciding whether the causal responsibility of a tuple for a Datalog query-answer is greater than a given threshold
  is {\em NP}-complete in data.

 \item We establish mutual characterizations between QA-causality and  different forms of delete-propagation as a view-update problem.

 \item We obtain that computing the size of the solution to a minimum-source-side-effect deletion-problem is hard for the complexity class $\nit{FP}^\nit{NP(log(n))}$, that of
 computational problems solvable in polynomial time (in data) by calling a logarithmic number of times an {\em NP}-oracle.

\item  We investigate in detail the problem of view-conditioned  QA-causality (vc-causality), and we establish connections with the view-side-effect free delete propagation problem for view updates.

 \item We obtain that deciding if an answer has a  {vc-}cause is {\em NP}-complete in data; that  deciding tuple  vc-causality is \nit{ {NP}}-complete in data; and
deciding if the  {vc-}causal responsibility of a tuple for a Datalog query-answer is greater than a given threshold
  is also {\em NP}-complete in data.

 \item We define the notion of QA-causality in the presence of integrity constraints (ICs), and investigate its properties. In particular, we make the case that the new property
 provides natural results.

   \item We obtain complexity results for QA-causality under ICs. In particular, we show that even for conjunctive queries, deciding tuple causality may become {\em NP}-hard under inclusion dependencies.

   \item We establish connections between QA-causality for Datalog queries under ICs and the view update problem and abduction from Datalog specifications, both under ICs. Through these connections
   we provide algorithmic results for computing causes for Datalog query answers under ICs.

\end{enumerate}

This paper is structured as follows. Section \ref{sec:prel} provides background material on relational databases and Datalog queries. Section \ref{sec:qa-causality} introduces the necessary concepts, known results, and
the main computational problems for QA-causality. Section \ref{sec:abdandcause} introduces the abduction problem in Datlog specifications, and establishes its connections with QA-causality. Section \ref{sec:delp&cause}
introduces the main problems related to updates trough views defined by monotone queries, and their connections with QA-causality problems. Section \ref{sec:vcc} defines and investigates view-conditioned QA-causality.
Section \ref{sec:c&ic} defines and investigates QA-causality under integrity constraints. Finally, Section \ref{sec:disc} discusses some relevant related problems and draws final conclusions. The Appendix contains a couple of proofs that are not in the main body of the paper.  \re{This paper is an extension of both \cite{uai15} and \cite{flairs16}.}

\section{Preliminaries}\label{sec:prel}

We  consider relational database schemas of the form $\mathcal{S} = (U, \mc{P})$, where $U$ is the possibly infinite
database domain and $\mc{P}$ is a finite set of {\em database predicates} of fixed arities.\footnote{ \ As opposed to built-in predicates, e.g. $\neq$, that we
leave implicit, unless otherwise stated.} A database instance $D$
compatible with $\mathcal{S}$ can be seen as a finite set of ground atomic formulas (a.k.a. atoms or tuples), of the form $P(c_1,..., c_n)$, where $P \in \mc{P}$ has arity $n$, and $c_1, \ldots, c_n \in U$.

A {\em conjunctive query}  ({CQ}) is a formula  of the first-order (FO) language $\mc{L}(\mc{S})$ associated to $\mc{S}$, of the form \ $\mc{Q}(\bar{x})\!: \exists \bar{y}(P_1(\bar{s}_1) \wedge \cdots \wedge P_m(\bar{s}_m))$,
where the $P_i(\bar{s}_i)$ are atomic formulas, i.e. $P_i \in \mc{P}$, and the $\bar{s}_i$ are sequences of terms, i.e. variables or constants of $U$. The $\bar{x}$ in  $\mc{Q}(\bar{x})$ shows
all the free variables in the formula, i.e. those not appearing in $\bar{y}$.  A sequence $\bar{c}$ of constants is an answer to query $\mc{Q}(\bar{x})$ if $D \models \mc{Q}[\bar{c}]$, i.e.
the query becomes true in $D$ when the free variables are replaced by the corresponding constants in $\bar{c}$. We denote the set of all answers from instance $D$ to a conjunctive query $\mc{Q}(\bar{x})$ with $\mc{Q}(D)$.

A conjunctive query is {\em Boolean} (a  {BCQ}), if $\bar{x}$ is empty, i.e. the query is a sentence, in which case, it is true or false
in $D$, denoted by $D \models \mc{Q}$ and $D \not\models \mc{Q}$, respectively. Accordingly, when $\mc{Q}$ is a  {BCQ}, $\mc{Q}(D) = \{\nit{yes}\}$ if $\mc{Q}$ is true, and $\mc{Q}(D) = \emptyset$, otherwise.

A query $\mc{Q}$ is {\em monotone} if for every two instances $D_1 \subseteq D_2$, \ $\mc{Q}(D_1) \subseteq \mc{Q}(D_2)$, i.e. the set of answers grows monotonically with the
instance. For example,  {CQ}s and unions of  {CQ}s ({UCQ}s) are monotone queries. In this work we consider only monotone queries.

An {\em integrity constraint} (IC) is a sentence $\varphi$ in the language $\mc{L}(\mc{S})$. For a given instance $D$ for schema $\mc{S}$, it may be true or false
in $D$, which is denoted with $D \models \varphi$, resp. $D \not \models \varphi$. Given a set $\Sigma$ of integrity constraints, a database instance $D$ is {\em consistent} if $D \models \Sigma$; otherwise it is said to be {\em inconsistent}.
In this work we assume that sets of integrity constraints are always finite and logically consistent (i.e. they are all simultaneously true in some instance).

A particular class of  ICs  is formed by {\em inclusion dependencies} ({IND}s), which are sentences of the form $\forall \bar{x} (P(\bar{x}) \rightarrow \exists \bar{y} R(\bar{x}',\bar{y}))$,
with $P,R$ predicates, $\bar{x}' \cap \bar{y} =\emptyset$, and $\bar{x}' \subseteq \bar{x}$. \re{ The  {\em tuple-generating dependencies} (tgds) are ICs that generalize INDs, and are
of the form \ $\forall \bar{x} (\bigwedge_i P_i(\bar{x}_i) \rightarrow \exists \bar{y} \bigwedge_j P_j(\bar{x}_j',\bar{y}_j))$, with $P_i, P_j$ predicates, $\bar{x}_j' \subseteq \bigcup \bar{x}_i = \bar{x}$, and
$\bar{y}_j \cap \bar{x} = \emptyset$.}

Another special class of ICs is formed by {\em functional dependencies} (FDs). For example,
 $\psi\!: \forall x \forall y \forall z (P(x, y) \land P(x, z) \rightarrow y = z)$ specifies that the second attribute of $P$ functionally depends upon the first. (If $A,B$ are the first and second attributes for $P$, the
 usual notation for this FD is \ $P\!: \ A \rightarrow B$.) Actually, this FD is also a  {\em key constraint} ({KC}),  in the sense that the attribute(s) on the LHS
of the arrow functionally determines all the other attributes of the predicate. \red{FDs form a particular class of {\em equality-generating dependencies} (egds), which are ICs of the form \ $\forall \bar{x} (\bigwedge_i P_i(\bar{x}_i) \rightarrow x_j = x_k))$, with $x_j,x_k \in \bar{x}$}
  (cf. \cite{Abiteboul95} for more details on ICs).

\re{Given a relational schema $\mc{S}$, queries $\mc{Q}_1(\bar{x}), \mc{Q}_2(\bar{x})$, and a set $\Sigma$ of ICs (all for schema
 schema $\mc{S}$), $\mc{Q}_1$ and $\mc{Q}_2$ {\em are equivalent wrt.} $\Sigma$, denoted $\mc{Q}_1 \equiv_{\Sigma} \mc{Q}_2$, iff $\mc{Q}_1(D) = \mc{Q}_2(D)$ for every instance
 $D$ for $\mc{S}$ that satisfies $\Sigma$.   One can define in similar terms the notion of query containment
 under ICs, denoted $\mc{Q}_1 \subseteq_{\Sigma} \mc{Q}_2$.}

A Datalog query $\mc{Q}(\bar{x})$ is a whole program $\Pi$ consisting of positive Horn rules (a.k.a.  positive definite rules), of the form \ $P(\bar{t}) \leftarrow P_1(\bar{t}_1), \ldots,
P_n(\bar{t}_n)$, with the $P_i(\bar{t}_i)$ atomic formulas. All the variables in $\bar{t}$ appear in some of the $\bar{t}_i$. Here,  $n \geq 0$, and if $n=0$, $P(\bar{t})$ is called a {\em fact} and does not contain variables. We assume the facts are those stored in an underlying extensional database $D$.

We may assume that a Datalog program $\Pi$ as a query  defines
an answer-collecting predicate $\nit{Ans}(\bar{x})$ by means of a top rule of the form $\nit{Ans}(\bar{t}) \leftarrow P_1(\bar{t}_1), \ldots, P_m(\bar{t}_m)$, where all the predicates
in the RHS (a.k.a. as the rule body) are defined by other rules in $\Pi$ or are database predicates for $D$. Here, the $\bar{t}, \bar{t}_i$ are lists of variables or constants, and the variables in $\bar{t}$ belong to
$\bigcup_i \bar{t}_i$.

Now, $\bar{a}$ is an answer to query $\Pi$ on $D$ when $\Pi \cup D \models \nit{Ans}(\bar{a})$. Here, entailment ($\models$) means that the RHS belongs to the minimal model of the LHS. So, the extension, $\nit{Ans}(\Pi\cup D)$, of predicate $\nit{Ans}$ contains the answers to the query in the minimal model of the program (including the database).
The Datalog query is Boolean if the top answer-predicate is propositional, with a  definition of the form $\nit{ans} \leftarrow P_1(\bar{s}_1), \ldots, P_m(\bar{s}_m)$. In this case, the query is true
if $\Pi \cup D \models \nit{ans}$, equivalently, if $\nit{ans}$ belongs to the minimal model of $\Pi \cup D$ \ \cite{Abiteboul95,ceri90}.

Datalog
queries may contain recursion, and then they may not be FO \cite{Abiteboul95,ceri90}. However they are also monotone.

\section{QA-Causality and its Decision Problems} \label{sec:qa-causality}

In this section we review the notion of QA-causality as introduced in \cite{Meliou2010a}. We also summarize the main decision and computational problems that emerge in this context and the established results for them.

\subsection{Causality and responsibility} \label{sec:causeintro}  In the rest of this work, unless otherwise stated, we assume that a relational database instance $D$ is split in two disjoint sets, $D=D^n \cup D^x$, where $D^n$ and $D^x$ are the sets of {\em endogenous} and {\em exogenous} tuples,
 respectively. 
\red{The former are tuples that we may consider as potential causes for data phenomena, tuples on which we have some form of control and can assess and modify. The latter are supposed to be given, unquestioned, and as such, not considered as possible causes. For example, they could be tuples provided by external sources we have no control upon.}

 A tuple $\tau \in D^n$ is  a
{\em counterfactual cause} for an answer $\bar{a}$ to $\mc{Q}(\bar{x})$ in $D$  if $D\models \mc{Q}(\bar{a})$, but $D\smallsetminus \{\tau\}  \not \models \mc{Q}(\bar{a})$.  A tuple $\tau \in D^n$ is an {\em actual cause} for  $\bar{a}$
if there  exists $\Gamma \subseteq D^n$, called a {\em contingency set}, such that $\tau$ is a counterfactual cause for $\bar{a}$ in $D\smallsetminus \Gamma$.  $\nit{Causes}(D, \mc{Q}(\bar{a}))$
denotes the set of actual causes for $\bar{a}$. If $\mc{Q}$ is Boolean, $\nit{Causes}(D, \mc{Q})$ contains the causes for answer $\nit{yes}$.
\re{For $\tau \in \nit{Causes}(D, \mc{Q}(\bar{a}))$, \ $\nit{Cont}(D, \mc{Q}(\bar{a}), \tau)$ denotes the set of contingency sets for $\tau$ as a cause for $\mc{Q}(\bar{a})$ in $D$.}

Notice that $\nit{Causes}(D, \mc{Q}(\bar{a}))$ is non-empty when $D \models \mc{Q}(\bar{a})$, but
$D^x  \not \models \mc{Q}(\bar{a})$, reflecting the fact that endogenous tuples are required for the answer.

Given a $\tau \in \nit{Causes}(D, \mc{Q}(\bar{a}))$, we collect all
subset-minimal  contingency sets associated with $\tau$:
\begin{eqnarray}
\re{\nit{Cont}^s(D, \mc{Q}(\bar{a}), \tau)}&:=&\{ \Gamma \subseteq D^n~|~ D\smallsetminus \Gamma \nonumber
\models Q(\bar{a}), ~D\smallsetminus (\Gamma \cup \{\tau\}) \not \models \mc{Q}(\bar{a}), \mbox{and }
\label{eq:ct}\\
   &&\!\!\!\hspace*{3.8cm}\forall \Gamma'\subsetneqq \Gamma, \ D \smallsetminus (\Gamma' \cup \{\tau\})
\models \mc{Q}(\bar{a}) \}. \nonumber
\end{eqnarray}
The {\em causal responsibility} of a tuple $\tau$ for answer $\bar{a}$, denoted with \re{$\rho_{_{\!\mc{Q}(\bar{a})\!}}^D(\tau)$}, is $\frac{1}{(|\Gamma| + 1)}$, where $|\Gamma|$ is the
size of the smallest contingency set for $\tau$. When $\tau$ is not an actual cause for $\bar{a}$, no contingency set is associated to $\tau$. In this case, $\rho_{_{\!\mc{Q}(\bar{a})\!}}^D(\tau)$ is defined as $0$. \red{In intuitive terms, the causal responsibility of a tuple $\tau$ is a {\em numerical measure} that is inversely proportional to the number of companion tuples that are needed to make $\tau$ a counterfactual cause.\footnote{\red{Non-numerical measures for the strengths of tuples as causes could be attempted, e.g. on the basis of minimality of contingency sets wrt. set inclusion, and then capturing a form of (more) specificity as a cause, but this is likely to produce  many incomparable causes. Under the responsibility degree, every two tuples can always be compared as causes.}} The less company $\tau$ needs to make the query true, the more responsibility it carries. This is the established notion of responsibility degree.\footnote{\red{However, in recent studies some objections have been raised in terms of how appropriately it  captures this intuition \cite{Braham09,halpResp,zultan,salimiThesis,tapp16}. Cf. \cite{tapp16} for a more detailed discussion, and the introduction of an alternative and also numerical measure, that of {\em causal effect} -so far for DBs without ICs- which appeals to auxiliary, uniform and independent probabilities associated to tuples, the notion of {\em lineage of a query} \cite{bunSig07,probDBs}, the expected value of a query as a Boolean variable, and the effect on it of modifying the tuple at hand.}}}

\red{We make note that ``causality for monotone queries" is monotonic, i.e. causes are never lost when new tuples are added to the database. However, for the same class of queries, ``most-responsible causality" is non-monotonic: the insertion of tuples into the database may make previous most responsible causes not such anymore (with other tuples taking this role). }

\vspace*{1mm}
\begin{example}\label{ex:cfex1}
Consider an instance $D$ with  relations  \nit{Author}({\nit{AName, JName}}) and \nit{Journal}({\nit{JName}, \nit{Topic}}, \nit{Paper\#}), and contents as
below:\\

{\small
\hspace*{0.5cm} \begin{tabular}{l|c|c|} \hline
\nit{Author} & \nit{AName} & \nit{JName} \\\hline
 & {\sf Joe}  & \sf{TKDE}\\
& {\sf John} & \sf{TKDE}\\
& \sf{Tom} & \sf{TKDE}\\
& {\sf John} & \sf{TODS}\\
 \hhline{~--} \end{tabular} ~~~~~~~\begin{tabular}{l|c|c|c|} \hline
\nit{Journal}  & \nit{JName} & \nit{Topic}& \nit{Paper\#} \\\hline
 & \sf{TKDE} & {\sf XML} & 30\\
& \sf{TKDE} & \sf{CUBE}& 31\\
& \sf{TODS} & {\sf XML} & 32\\
 \hhline{~---}
\end{tabular}
} \\ \\
The conjunctive query:
\begin{eqnarray}
\mc{Q}(\nit{AName}, \nit{Topic})\!:&& \!\!\!\! \!\! \exists \nit{J\!Name} \ \exists \nit{Paper\#}(\nit{Author}(\nit{AName, J\!Name})  \wedge \label{eq:query}\\
&&\hspace*{2cm}\nit{Journal}(\nit{J\!Name, Topic}, \nit{Paper\#})) \nonumber
\end{eqnarray}
has the following answers:

\vspace{-3mm}
\begin{multicols}{2}
{\small
\begin{tabular}{l|c|c|} \hline
$\mc{Q}(D)$ & \nit{AName} & \nit{ \ Topic \ }\\\hline
 & {\sf Joe}  & {\sf XML}\\
  & {\sf Joe}  & \sf{CUBE}\\
& \sf{Tom} & {\sf XML}\\
& \sf{Tom} & \sf{CUBE}\\
& {\sf John} & {\sf XML}\\
& {\sf John} & \sf{CUBE}\\
 \hhline{~--}
 \end{tabular}
}

\vspace{3mm}
\noindent Assume   $\langle \nit{\sf John},{\sf XML} \rangle$ is an unexpected answer to $\mc{Q}$. That is, it is not likely that {\sf John} has a paper on {\sf XML}. Now, we want to compute causes for this unexpected observation. For the moment assume  all tuples in $D$ are endogenous.

\end{multicols}

It holds that \nit{Author({\sf John}, {\sf TODS})} is an actual cause for answer $\langle \nit{\sf John},{\sf XML} \rangle$. Actually,
it has two contingency sets, namely: $\Gamma_1 = \{\nit{Author(\mbox{\sf John},\mbox{\sf TKDE})}\}$
 and $\Gamma_2$=\{\red{\nit{Journal}(\mbox{\sf TKDE},\mbox{\sf XML},\mbox{30})}\}. That is, \nit{Author({\sf John},{\sf TODS})} is a counterfactual
 cause for
  $\langle \nit{\sf John},{\sf XML}\rangle$ in both $D \smallsetminus \Gamma_1$ and $D \smallsetminus \Gamma_2$.  Moreover, the responsibility of  \nit{Author({\sf John},{\sf TODS})}  is $\frac{1}{2}$, because its minimum-cardinality contingency sets have size 1.

Tuples  \red{\nit{Journal}(\mbox{\sf TKDE},\mbox{\sf XML},\mbox{30})}, \nit{Author(\mbox{\sf John},\mbox{\sf TKDE})} and  \nit{Journal}({\sf TODS}, {\sf XML}, \mbox{32})  are also actual causes  for $\langle \nit{\sf John},{\sf XML} \rangle$, with responsibility $\frac{1}{2}$.

For more subtle situation, assume only \nit{Author} tuples are endogenous, possibly reflecting
 the fact that the data in \nit{Journal} table are more reliable than those in the \nit{Author} table. Under this assumption,
 the only actual causes for answer
  $\langle \nit{\sf John},{\sf XML} \rangle$  are  \nit{Author(\mbox{\sf John},\mbox{\sf TKDE})}
and \nit{Author}(\nit{\sf John},{\sf TODS}).
\boxtheorem
\end{example}

The definition of QA-causality can be applied without any conceptual changes to Datalog queries.  Actually,
 {CQ}s can be expressed as Datalog queries. For example, (\ref{eq:query}) can be expressed in Datalog as: \

$\nit{Ans}_{\mc{Q}}(\nit{AName}, \nit{Topic}) \leftarrow \nit{Author}(\nit{AName, JName}),$

\hspace*{6.5cm}$\nit{Journal}(\nit{JName, Topic}, \nit{Paper\#})$,\\
with the auxiliary predicate $\nit{Ans}_{\mc{Q}}$ collecting the answers to query $\mc{Q}$.

In the case of Datalog, we sometimes use the notation $\nit{Causes}(D, \Pi(\bar{a}))$ for the set of causes
for answer $\bar{a}$ (and simply $\nit{Causes}(D, \Pi)$ when $\Pi$ is Boolean).

\begin{example} \em \label{ex:datalogcause} \em
Consider the instance $D$ with a single binary relation $E$ as below ($t_1$-$t_7$ are tuple identifiers). Assume all tuples are endogenous.

Instance $D$ can be represented as the directed graph $G(\mc{V}, \mc{E})$ in Figure \ref{fig:gresp}, where the set of vertices $\mc{V}$ coincides with the active domain of $D$ (i.e. the set of constants in $E$). The set
 of edges $\mc{E}$ contains $(v_1, v_2)$ iff $E(v_1, v_2) \in D$.  The tuple identifiers are used as labels for the corresponding edges, and also to refer to the database tuples.

\begin{figure}
  \centering
\includegraphics[width=2in]{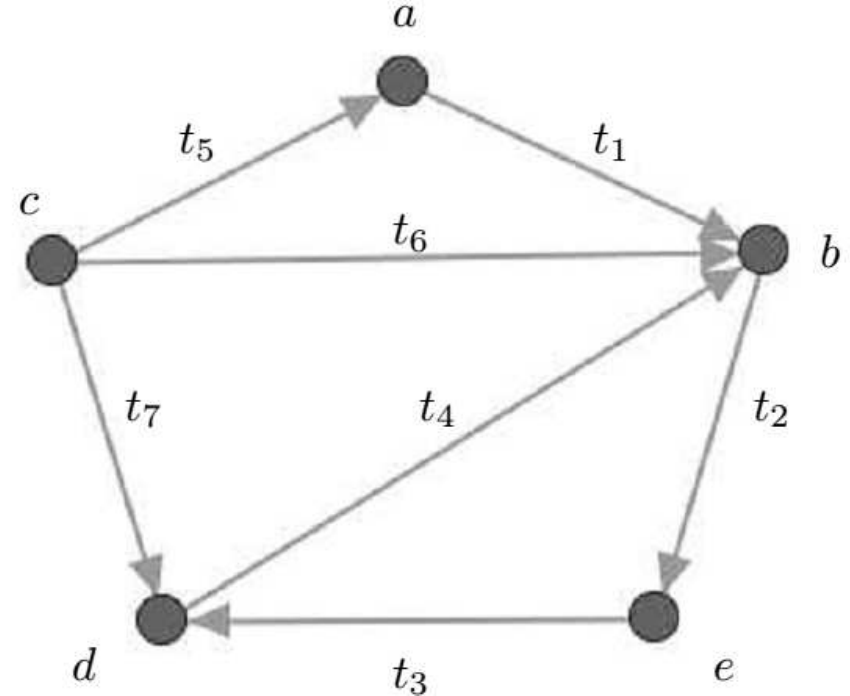}
  \caption{Graph representing a database }\label{fig:gresp}
\end{figure}

\begin{multicols}{2}

\begin{center} \begin{tabular}{l|c|c|} \hline
$E$  & A & B \\\hline
$t_1$ & {\sf a} & $b$\\
$t_2$& $b$ & $e$\\
$t_3$& $e$ & $d$\\
$t_4$& $d$ & $b$\\
$t_5$& $c$ & {\sf a}\\
$t_6$& $c$ & $b$\\
$t_7$& $c$ & $d$\\ \cline{2-3}

\end{tabular}
\end{center}

\noindent Consider the recursive Datalog query $\Pi$: 
\begin{eqnarray}
       \nit{Ans}(x, y) &\leftarrow& P(x, y) \nonumber \\
       P(x, y)&\leftarrow& E(x, y) \nonumber \\
       P(x, y)&\leftarrow& P(x, z), E(z, y),\nonumber
\end{eqnarray}
which collects pairs of vertices of $G$ that are connected through a path.

\end{multicols}

 Since \ $\Pi \cup D \models \nit{Ans}(c, e)$, we have $\langle c, e \rangle$ as an answer to query $\Pi$ on $D$.  This is because there are three distinct paths between
 $c$ and $e$ in $G$. All tuples except for $t_3$ are actual causes for this
 answer: \ $\nit{Causes}(E, \Pi(c, e))=\{t_1, t_2, t_4, t_5, t_6, t_7\}$. We can see that
  all of these tuples contribute to at least one path between $c$ and $e$. Among them, $t_2$ has the
   highest responsibility, because, $t_2$ is a counterfactual cause for the answer, i.e. it has an empty contingency set.
\boxtheorem
\end{example}

\ignore{
\psscalebox{1.0 1.0} 
{\begin{pspicture}(0, -2.4845312)(6.18, 2.4845312)
\rput(2.93, 0.08569336){\includegraphics{GRAPH.eps}}
\rput[bl](2.8, 2.5856933){$a$}
\rput[bl](6.0, 0.78569335){$b$}
\rput[bl](5.2, -2.3043066){$e$}
\rput[bl](0.4, -2.3043066){$d$}
\rput[bl](0.0, 1.1856934){$c$}
\rput[bl](4.3, 1.5856934){$t_1$}
\rput[bl](1.2, 1.5856934){$t_5$}
\rput[bl](2.8, 0.90569335){$t_6$}
\rput[bl](0.8, -0.41430664){$t_7$}
\rput[bl](3.0, -0.41430664){$t_4$}
\rput[bl](2.8, -2.4143066){$t_3$}
\rput[bl](5.5, -0.41430664){$t_2$}
\end{pspicture}
}
}



The complexity of the computational and decision problems that arise in QA-causality  have been investigated in \cite{Meliou2010a, icdt15}. Here we recall those results that we will
use throughout this work. The first problem is about deciding whether a tuple is an actual cause for a query answer.
\begin{definition} \em   \label{def:cp}    For a  Boolean monotone query $\mc{Q}$,
the {\em causality decision problem} (CDP) is (deciding about membership of):

$\mc{CDP}(\mc{Q}) := \{(D, \tau)~|~ \tau \in D^n, \mbox{and } \tau \in  \nit{Causes}(D, \mc{Q}) \}.$ \boxtheorem
\end{definition}

\red{This problem is tractable for \nit{UCQ}s \cite{Meliou2010a,tocs15}, because it can be solved by CQ answering in relational databases.} The next problem is about deciding if the responsibility of a tuple as a cause for a query answer
is above a given
 threshold.
\begin{definition} \em   \label{def:resp}
   For a  Boolean monotone query $\mc{Q}$,
the {\em responsibility decision problem} (RDP) is (deciding about membership of):

 $\mathcal{RDP}(\mc{Q})=\{(D, \tau, v)~|~ \tau \in D^n, v \in \{0\} \ \cup $\\
 \hspace*{4.7cm} $\{\frac{1}{k}~|~k \in \mathbb{N}^+\}, $  $D \models \mc{Q}$ \ and \ $\rho_{_{\!\mc{Q}\!}}^D(\tau) > v  \}$. \boxtheorem
\end{definition}

This problem is \nit{ {NP}}-complete for CQs \cite{Meliou2010a} and {UCQ}s \cite{icdt15}, but tractable  for {\em linear}  {CQ}s \cite{Meliou2010a}. Roughly speaking, a  {CQ}
is linear if its atoms can be ordered in a way that every variable appears in a continuous sequence of atoms that does not contain a self-join (i.e. a join involving the same predicate), e.g. $\exists xvyu(A(x) \wedge S_1(x, v) \wedge S_2(v, y) \wedge R(y, u) \wedge S_3(y, z))$ is linear, but not $\exists x y z(A(x) \wedge B(y) \wedge C(z) \wedge W(x, y, z))$, for which  {RDP} is \nit{ {NP}}-complete \cite{Meliou2010a}.  \ignore{The class of  {CQ}s for which   {RDP} is tractable can be extended  to {\em weakly linear} \cite{Meliou2010a}.\footnote{ \ Computing  sizes of minimum contingency sets for this class is reduced to the max-flow/min-cut problem in a network.}}

\ignore{\comlb{Check previous paragraph. Does it still hold? It was related to the wrong dichotomy result, but may still be true and relevant. Ref. for the notion and claim about weakly-linear? Should we get rid of all the stuff
on linear queries?}}

The functional, non-decision, version of  {RDP} is about computing
responsibilities. This optimization problem is complete (in data) for $\nit{FP}^{\nit{ {NP}(log} (n))}$ for  {UCQ}s \cite{icdt15}.
Finally, we have the problem of deciding whether a tuple is a most responsible cause:
\begin{definition} \em   \label{def:mracp}   For a  Boolean monotone query $\mc{Q}$, the {\em most responsible cause decision problem} \ignore{(\re{{MRDP}})}
 is:

   $\mc{MRCD}(\mc{Q})$ $=\{(D, \tau)~|~ \tau \in D^n \ \mbox{and } 0 < \rho_\mc{Q}^D(\tau) \mbox{ is a maximum for } D\}$.
\boxtheorem
\end{definition}


For  {UCQ}s this problem is complete for  $\nit{P}^{\nit{ {NP}(log} (n))}$ \cite{icdt15}. \re{Hardness already holds for a CQ. }

A notion of {\em view-conditioned causality}  \cite{Meliou2010b}   will be formalized and investigated in Section \ref{sec:vcc}.

\section{Causality and Abduction} \label{sec:abdandcause}

In general  logical terms, an abductive explanation for an observation is a formula that, together with a background logical theory, entails the observation. Although one could see an abductive explanation as
a cause for the observation, it has been argued that causes and abductive explanations are not necessarily the same \cite{Psillos96, DeneckerK02}. \ignore{A classic example from \cite{Psillos96} is as follows:  the disease
{\em paresis} is caused by a latent untreated form of {\em syphilis}. The probability that latent untreated syphilis leads
to paresis is only 25\%. Therefore, syphilis is the cause of paresis but does not entail it, while
paresis entails syphilis but does not cause it.}

Under the abductive approach to diagnosis \cite{console91, EiterGL95, Poole92, Poole94}, it is common that the system specification rather explicitly describes causal information,
specially in action theories where the
effects of actions are directly represented by positive definite rules\ignore{Horn formulas}. By restricting the explanation formulas to the predicates describing primitive
causes (action executions), an explanation formula which entails an observation gives also a cause for the observation \cite{DeneckerK02}. In this
case, and is some sense, causality information is imposed by the system specifier \ \cite{Poole92}.

In database causality we do not have, at least not initially, a system description,\footnote{ \ Having integrity constraints would go in that direction, but this is something that has not been
considered in database causality so far. See \cite[sec. 5]{icdt15} for a consistency-based diagnosis connection, where the DB is turned into a theory.} but just a set of tuples. It is when we pose a query that we create something like a description, and the causal relationships between tuples are captured
by the combination of  atoms in the query. If the query is a Datalog query (in particular, a  {CQ}), we have a specification in terms of positive definite rules.

 In this section we will first establish
 connections between abductive diagnosis and database causality.\footnote{ \ In \cite{icdt15} we established such a connection
 between another form of model-based diagnosis \cite{struss}, namely consistency-based diagnosis \cite{Reiter87}. For relationships
and comparisons between consistency-based and abductive diagnosis see \cite{console91}.} We start by making  precise  the kind of abduction problems we will consider.

\subsection{Background on Datalog abductive diagnosis}\label{sec:backAbd}

A  {\em Datalog abduction problem} \cite{EiterGL97} is of the form $\mathcal{AP}= \langle \Pi, E, \nit{Hyp}, \nit{Obs}\rangle$, where: \ (a) $\Pi$ is a set of Datalog rules,
 (b) $E$ is a  set of ground atoms (the extensional database), (c) $\nit{Hyp}$, the hypothesis, is a finite set of
ground atoms, \ignore{{whose predicates do not appear in heads of the rules in $\Pi$},} the abducible atoms in this case,\footnote{ \ It is common to accept as hypothesis all the possible ground instantiations of {\em abducible predicates}. We assume abducible
predicates do not appear in rule heads.} and (d) $\nit{Obs}$, the observation, is a finite conjunction of
ground atoms. As it is common, we will start with the assumption that $\Pi \cup E \cup \nit{Hyp} \models \nit{Obs}$. $\Pi \cup E$ is called the {\em background theory} (or specification).

\begin{definition} \label{def:DAP} \em  Consider  a {\em Datalog abduction problem} $\mathcal{AP}= \langle \Pi, E, \nit{Hyp}, \nit{Obs}\rangle$.\vspace{-2mm}
\begin{enumerate}[(a)] \item
 An  {\em  abductive diagnosis} (or simply, {\em a solution}) for $\mathcal{AP}$ is a subset-minimal $\Delta \subseteq \nit{Hyp}$, such that
$ \Pi \cup E \cup \Delta \models \nit{Obs}$.\footnote{\red{The minimality requirement is common in model-based diagnosis, so as in many non-monotonic reasoning tasks in knowledge representation. In particular, its use in this work is not due to the use of Datalog, for which the minimal-model semantics is adopted.}}

This requires that no proper subset of $\Delta$ has this property.\footnote{ \ Of course, other minimality criteria could take this place.}
 $\nit{Sol}(\mathcal{AP})$ denotes the set of abductive diagnoses for problem $\mc{AP}$.
 \item
A hypothesis $h \in \nit{Hyp}$  is {\em relevant} for $\mathcal{AP}$ if  $h$ \red{is} contained in at least one diagnosis of $\mathcal{AP}$, otherwise it is {\em irrelevant}. $\nit{Rel}(\mc{AP})$ collects  all relevant hypothesis for $\mc{AP}$.
\item A hypothesis $h \in \nit{Hyp}$  is {\em necessary} for $\mathcal{AP}$ if  $h$ \red{is} contained in all diagnosis of $\mathcal{AP}$. $\nit{Ness}(\mc{AP})$ collects  all the necessary hypothesis for $\mc{AP}$.
\boxtheorem
\end{enumerate}
\end{definition}
Notice that for a problem $\mathcal{AP}$, $\nit{Sol}(\mathcal{AP})$ is never empty due to the assumption $\Pi \cup D \cup \nit{Hyp} \models \nit{Obs}$. In case, $\Pi \cup D \models \nit{Obs}$, it holds $\nit{Sol}(\mathcal{AP}) = \{\emptyset\}$.

\begin{figure}
  \centering
    \includegraphics[width=2.5in]{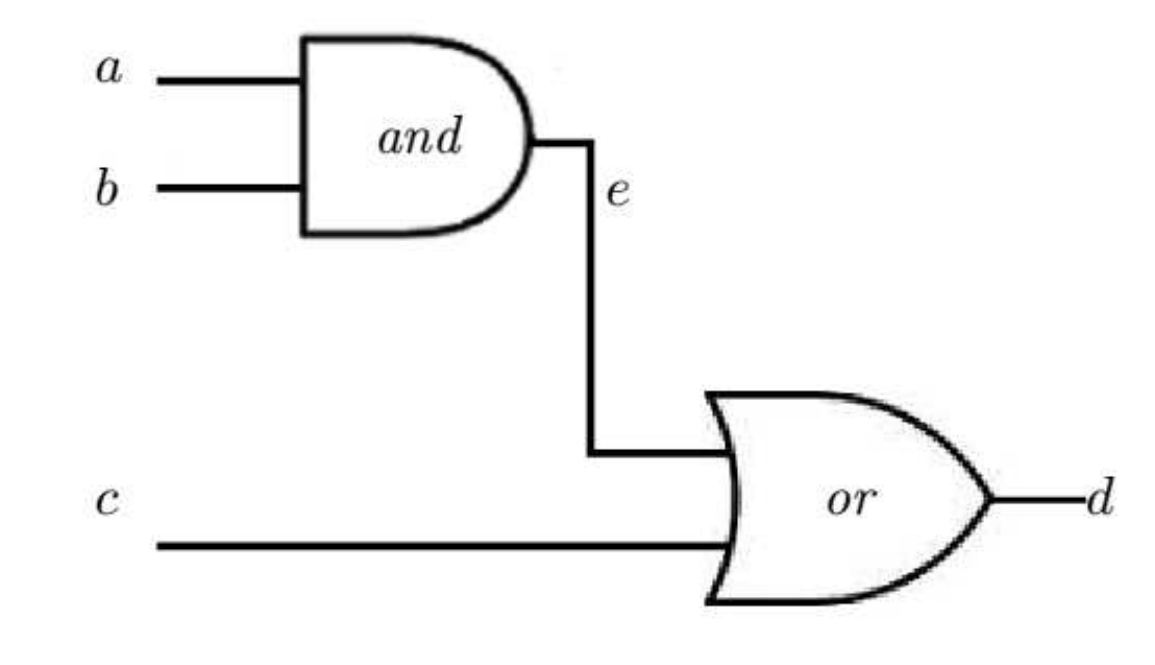}
  \caption{A simple circuite with two gates}\label{fig:cir}
\end{figure}

\begin{example} \em \label{ex:datalogabduction} \em
Consider the  digital circuit in Figure \ref{fig:cir}. The inputs are $a = 1$, $b = 0$, $c = 1$, but the output is $d = 0$. So, the circuit is not working properly. The
diagnosis problem is formulated below as a Datalog abduction problem whose data domain is $\{{\sf a},{\sf b},{\sf c},{\sf d},{\sf e},{\sf and}, {\sf or}\}$.
The underlying,  extensional database is as follows:  $E=\{\nit{One}({\sf a}), \nit{Zero}({\sf b}),$ $\nit{One}({\sf c}), \nit{And}({\sf a}, {\sf b}, {\sf e}, {\sf and}), \nit{Or}({\sf e}, {\sf c}, {\sf d}, {\sf or}\}$.

The Datalog program $\Pi$ contains rules that model
the normal and the faulty behavior of each gate. We show only the Datalog rules for the
\nit{And} gate.  For
its normal behavior, we have the following rules:

\vspace{-.5cm}
\begin{eqnarray}
       \nit{One}(O) &\leftarrow& \nit{And}(I_1, I_2, O, G), \nit{One}(I_1), \nit{One}(I_2) \nonumber \\
       \nit{Zero}(O) &\leftarrow& \nit{And}(I_1, I_2, O, G), \nit{One}(I_1), \nit{Zero}(I_2) \nonumber \\
       \nit{Zero}(O) &\leftarrow& \nit{And}(I_1, I_2, O, G), \nit{Zero}(I_1), \nit{One}(I_2) \nonumber \\
       \nit{Zero}(O) &\leftarrow& \nit{And}(I_1, I_2, O, G), \nit{Zero}(I_1), \nit{One}(I_2). \nonumber
\end{eqnarray}
 The
faulty behavior is modeled by the following
rules:
\begin{eqnarray}
       \nit{Zero}(O) &\leftarrow& \nit{And}(I_1, I_2, O, G), \nit{One}(I_1), \nit{One}(I_2), \nit{Faulty}(G) \nonumber \\
       \nit{One}(O) &\leftarrow& \nit{And}(I_1, I_2, O, G), \nit{One}(I_1), \nit{Zero}(I_2), \nit{Faulty}(G) \nonumber \\
       \nit{One}(O) &\leftarrow& \nit{And}(I_1, I_2, O, G), \nit{Zero}(I_1), \nit{One}(I_2), \nit{Faulty}(G) \nonumber \\
       \nit{One}(O) &\leftarrow& \nit{And}(I_1, I_2, O, G), \nit{Zero}(I_1), \nit{One}(I_2), \nit{Faulty}(G) \nonumber
\end{eqnarray}
Finally, we consider $\nit{Obs}\!: \nit{Zero}({\sf d})$, and
$\nit{Hyp}=\{\nit{Faulty({\sf and})}, \nit{Faulty({\sf or})}\}$. The abduction problem consists in finding minimal $\Delta \subseteq \nit{Hyp}$, such that
$\Pi \cup E \cup \Delta \models \nit{Zero}({\sf d})$. There is one abductive diagnosis: $\Delta=\{\nit{Faulty({\sf or})}\}$. \boxtheorem
\end{example}

\ignore{The following problems naturally arise in the context of
Datalog abduction: a) Relevance problem: does a hypothesis $h$ contribute to some solution of $\mc{AP}$?; b) Necessity problem:
does a hypothesis $h$ occur in all solutions of $\mc{AP}$? More precisely, we consider the following decision problem. }

In the context of Datalog abduction, we are interested in deciding, for a fixed Datalog program, if a hypothesis is relevant/necessary or not, with all the data as input.
More precisely, we consider the following decision problems.

\begin{definition} \em   \label{def:relp}  Given a Datalog program $\Pi$,

\noindent (a) The {\em necessity decision problem} ({NDP}) {\em for} $\Pi$ is \ (deciding about the membership of): 

\vspace{1mm}
\hspace*{-6mm}$\mc{NDP}(\Pi)=\{(\nit{E, Hyp, Obs}, h)~|~ h \in \nit{Ness}(\mc{AP}), \mbox{with } 
\mc{AP}= \langle \Pi, E, \nit{Hyp}, \nit{Obs}\rangle \ignore{\mbox{and } h \in \nit{Hyp}} \}$.

\noindent (b) The {\em relevance decision problem}  ({RLDP}) {\em for}  $\Pi$ is \ (deciding about the membership of):

\vspace{1mm}
\hspace*{-6mm}$\mc{RLDP}(\Pi)=\{(\nit{E, Hyp, Obs}, h)~|~ h \in \nit{Rel}(\mc{AP}), \mbox{with }
\mc{AP}= \langle \Pi, E, \nit{Hyp}, \nit{Obs}\rangle \ignore{\mbox{ and } h \in \nit{Hyp} }\}$.
\boxtheorem
\end{definition}
As it is common, we will assume that $|\nit{Obs}|$, i.e. the number of atoms in the conjunction,
is bounded above by a fixed parameter $p$. \ignore{It is common that} In many cases, $p=1$ (a single atomic
observation). 


The last two definitions \ignore{ \ref{def:relp}} suggest that we are interested in the {\em data complexity} of the relevance and necessity decision problems for Datalog abduction. That is, the Datalog program is fixed,
but the data consisting of
hypotheses and input structure $E$ may change. In contrast, under {\em combined complexity} the program is also part of the input, and the complexity is measured also
in terms of the program size.

A comprehensive complexity analysis of several reasoning tasks on abduction from propositional logic programs, in particular of the relevance and necessity problems,  can be found in \cite{EiterGL97}. Those results are all in combined complexity. In  \cite{EiterGL97}, it has been shown that for abduction from function-free first-order logic programs, the data complexity of each type of reasoning problem in
the first-order case coincides with \re{the \ignore{\red{combined}} complexity} of the same type of reasoning
problem in the propositional case. In this way, the next two results can be obtained for {NDP} and {RLDP} from \cite[theo. 26]{EiterGL97} and \re{the \ignore{\re{combined}} complexity} of these problems for {\em propositional Horn abduction} (PDA), established in \cite{Friedrich90} \red{(cf. also  \cite{EiterGL95})}. \re{In the Appendix we provide direct,  {\em ad hoc}  proofs by  adapting the full machinery developed in \cite{EiterGL97} for  general programs.} \red{The next result follows from the membership  of \nit{PTIME} in data complexity of Datalog query evaluation (actually, this latter problem is \nit{PTIME}-complete in data \cite{dantsin}).}

\ignore{\comlb{Again on this, in the parts on complexity in red right above we used to have ``combined" for the propositional case. The point is that in the propositional case we do not talk about ``combined" complexity, because
there is no clear separation between the data and the formulas. So, I removed the ``combined" that we used to have.}}


\begin{proposition}\label{pro:nessp} \em For every Datalog program, $\Pi$,
 $\mc{NDP}(\Pi)$ is in  {\nit{PTIME}} (in data).
 \boxtheorem
\end{proposition}

\begin{proposition}\label{pro:relp} \em For Datalog programs $\Pi$, $\mc{RLDP}(\Pi)$
 is {\em  {NP}}-complete \ (in \linebreak data).\footnote{ \ More precisely, this statement (and others of this kind) means: (a) For every Datalog program $\Pi$,
$\mc{RLDP}(\Pi) \in \nit{ {NP}}$; and
(b) there are programs $\Pi'$ for which $\mc{RLDP}(\Pi')$
 is {\em  {NP}}-hard \ (all this in  data).}\boxtheorem
\end{proposition}
It is clear from this result that  deciding relevance for Datalog abduction is also intractable in combined complexity.  However, a tractable case of combined complexity is identified  in \cite{Gottlob10b}, on the basis of the notions of {\em tree-decomposition and bounded tree-width}, which we now briefly present.

Let $\mc{H} = \langle V, H \rangle$ be a hypergraph, where  $V$ is the set of vertices, and $H$ \bblue{is} the set of hyperedges,
 i.e. of subsets of $V$. A tree-decomposition  of $\mc{H}$ is a pair $(\mc{T}, \lambda)$, where $\mc{T} = \langle N, E\rangle$ is a tree and $\lambda$ is a labeling
function that assigns to each node $n \in N$, a subset  $\lambda(n)$ of  $V$ ($\lambda(n)$ is aka. bag), i.e. $\lambda(n) \subseteq V$, such that, for every node $n \in N$, the following hold: \ (a) For every $v \in V$, there exists $n \in N$ with
$v \in \lambda(n)$. \
(b) For every $h \in H$, there exists a node $n \in N$ with $h \subseteq \lambda(n)$. \ (c) For every $v \in V$, the set of nodes $\{n ~|~ v \in \lambda(n)\}$ induces a connected subtree of $\mc{T}$.

The {\em width} of a tree decomposition $(\mc{T}, \lambda)$ of $\mc{H} = \langle V, H \rangle$, with $\mc{T} = \langle N, E\rangle$, is defined as $\nit{max}\{|\lambda(n)| - 1 \ : \  n \in N\}$.  The tree-width $t_w(\mc{H})$ of $\mc{H}$ is
the minimum width over all its tree decompositions.

Intuitively, the tree-width of a hypergraph $\mc{H}$ is a
measure of the ``tree-likeness" of $\mc{H}$. A set of vertices that form a cycle in $\mc{H}$ are put into a same bag, which becomes (the bag of a) node in the corresponding tree-decomposition.
If the tree-width of the
hypergraph under consideration is bounded by a fixed constant,
then many otherwise intractable problems become tractable \cite{Gottlob10}.

It is possible to associate an hypergraph to any finite structure $D$ (think of a relational database): If its universe (the active domain in the case of a relational database) is $V$, define the hypergraph  $\mc{H}(D) = (V, H)$, with
$H = \{~\{a_1, \ldots, a_n\}~ | ~ D$ contains a ground atom $P(a_1 \ldots a_n)$ for some predicate symbol $P \}$.

\begin{figure}[h!]
  \centering
  \includegraphics[width=4.5in]{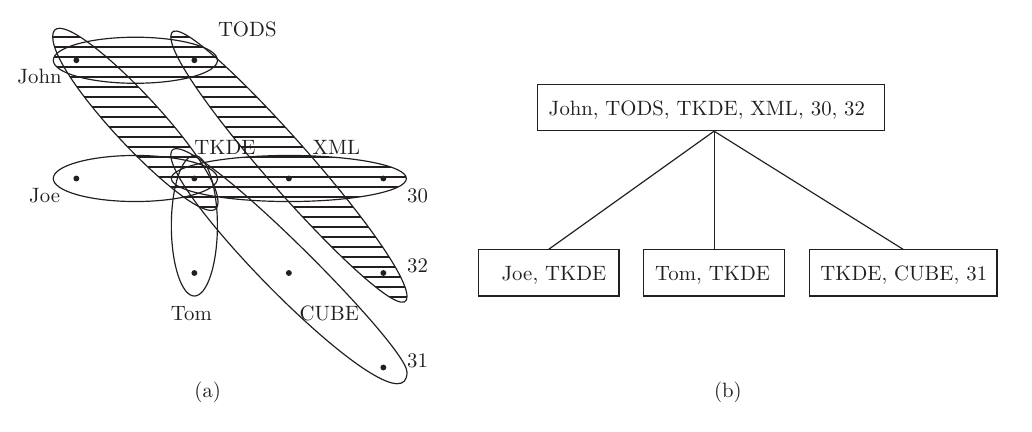}
  \vspace{-5mm}
  \caption{\ (a) \ $\mc{H}(D)$. \  \ (b) \ A  tree decomposition of $\mc{H}(D)$. }\label{fig:fig1}
\end{figure}
\begin{example}\label{ex:cfex3} Consider instance $D$ in Example \ref{ex:cfex1}. The hypergraph $\mc{H}(D)$ associated to $D$ is shown in Figure \ref{fig:fig1}(a). Its vertices are the elements of
$\nit{Adom}(D) = \{{\sf John}, {\sf Joe}, {\sf Tom}, {\sf TODS}, {\sf TKDE}, {\sf XML}, {\sf CUBE}, \mbox{ 30, 31, 32}\}$, the active domain of $D$. For example, since
$\nit{Journal}({\sf TKDE}, {\sf XML},\mbox{30}) \in D$, $\{{\sf TKDE}, {\sf XML},\mbox{30}\}$ is one of the hyperedges.

The dashed ovals show four sets of vertices, i.e. hyperedges, that together form a cycle. Their elements are put into the same bag of the tree-decomposition. Figure \ref{fig:fig1}(b) shows a possible tree-decomposition of $\mc{H}(D)$.
In it, the maximum $|\lambda(n)| -1$ is $6 -1$, corresponding to the top box bag of the tree. So, $t_w(\mc{H}(D)) \leq 5$.\boxtheorem
\end{example}
The following is a {\em fixed-parameter tractability} result for the relevance decision problem for Datalog abduction \ignore{problems}
for {\em guarded} programs $\Pi$, where in every rule body there
is an atom that contains (guards) all the variables appearing in that body.

\begin{theorem} \label{the:trac} \em  \cite[theo. 7.9]{Gottlob10b} \
Let $k$ be an integer. For Datalog abduction problems $\mathcal{AP}=\langle \Pi, E,
\nit{Hyp}, \nit{Obs}\rangle$ where $\Pi$ is guarded, and
$t_w(\mc{H}(E)) \leq k$, relevance can be decided in polynomial time in $|\mc{AP}|$.\ignore{\footnote{ \ This is Theorem 7.9 in \cite{Gottlob10b}.}} More precisely, the following decision problem is tractable:

$\mc{RLDP}=\{(\langle \Pi, \nit{E, Hyp, Obs}\rangle, h)~|~ h \in \nit{Rel}(\langle \Pi, \nit{E, Hyp, Obs}\rangle),
 h \in \nit{Hyp}, $\\
 \hspace*{6.1cm} $\Pi \mbox{ is guarded, and } t_w(\mc{H}(E)) \leq k \}$.
\boxtheorem
\end{theorem}
This is a case of tractable combined complexity with a  fixed parameter that is the
tree-width of the extensional database.

\re{\bf \em In the rest of this  \re{section} we assume, unless otherwise stated,  that we have a  partitioned relational instance  $D=D^x \cup D^n$.}

\subsection{Actual causes from abductive diagnoses} \label{sec:abddig}

In this section we show that, for  Datalog system specifications, abductive inference
corresponds to actual causation. 
That is, abductive diagnoses for an observation essentially contain actual causes for the observation.

Consider that $\Pi$ is a Boolean, possibly recursive Datalog query; and assume that $\Pi \cup D \models \nit{ans}$. Then, the decision problem in Definition \ref{def:cp} takes the form:
\begin{equation}\label{eq:causalDatalog}
\mc{CDP}(\Pi) := \{(D, \tau)~|~ \tau \in D^n, \mbox{and } \tau \in  \nit{Causes}(D, \Pi) \}.
\end{equation} \ignore{The  {RDP} in Definition \ref{def:resp}, can be adopted for Datalog queries
 in a similar way.}
We now show that  actual causes for $\nit{ans}$ can be obtained from abductive diagnoses of the associated {\em causal Datalog abduction problem} ({CDAP}): \  $\mathcal{AP}^c:=\langle \Pi, D^x, D^n, \nit{ans}\rangle$, where
 $D^x$ takes the role of the extensional database for $\Pi$. Accordingly, $\Pi \cup D^x$ becomes  the {\em background theory}, $D^n$ becomes the set of {\em hypothesis}, and atom $\nit{ans}$ is the observation.

\begin{proposition} \label{pro:abdf&cfcaus}  \em
For an instance $D=D^x \cup D^n$ and a Boolean Datalog query $\Pi$, with $\Pi \cup D \models \nit{ans}$, and   its associated CDAP $\mc{AP}^c$, the following hold:

\begin{enumerate}[(a)]
\item   $ \tau \in D^n$ is an counterfactual cause for $\nit{ans}$  iff $\tau \in \nit{Ness}(\mathcal{AP}^c)$.

\item  $\tau \in D^n$ is an actual cause for $\nit{ans}$  iff $\tau \in \nit{Rel}(\mathcal{AP}^c)$. 
\end{enumerate}
\end{proposition}

\dproof{Part (a) is straightforward. To proof part (b), first assume $\tau$ is an actual cause for
\nit{ans}. According to the definition of an actual cause, there
 exists a contingency set  $ \Gamma \subseteq D^n$ such that $ \Pi \cup D \smallsetminus
  \Gamma \models  \nit{ans}$ but $\Pi \cup D \smallsetminus (\Gamma \cup \{\tau\}) \not \models  \nit{ans}$. This implies
  that there exists a set $\Delta \subseteq D^n$ with $\tau \in \Delta $ such that   $ \Pi \cup \Delta \models \nit{ans}$.
   It is easy to see that  $ \Delta$ is an abductive diagnosis for $\mathcal{AP}^c$. Therefore, $\tau \in \nit{Rel}(\mathcal{AP}^c)$.

Second, assume $\tau \in \nit{Rel}(\mathcal{AP}^c)$. Then there exists a set
 $\mc{S}_k \in \nit{Sol}(\mathcal{AP}^c)=\{s_1 \ldots s_n\}$ such that
 $\mc{S}_k \models \nit{ans}$ with $\tau \in \mc{S}_k$. Obviously, $\nit{Sol}(\mathcal{AP}^c)$ is a
 collection of subsets of $D^n$. Pick a  set $\Gamma \subseteq D^n$ such that for all
  $\mc{S}_i \in \nit{Sol}(\mathcal{AP}^c)$ $i \not = k$, $\Gamma \cap \mc{S}_i \not = \emptyset$ and
    $\Gamma \cap \mc{S}_k =\emptyset$. It is clear that $ \Pi \cup D \smallsetminus (\Gamma \cup \{t\}) \not \models \nit{ans}$ but
     $\Pi \cup D \smallsetminus \Gamma \models \nit{ans}$. Therefore, $\tau$ is an actual cause for \nit{ans}.
To complete the proof we need to show that such $\Gamma$ always exists. This
 can be done by applying
the digitalization technique to construct such $\Gamma$. Since all elements of $\nit{Sol}(\mathcal{AP}^c)$
are subset-minimal, then, for each $\mc{S}_i \in \nit{Sol}(\mathcal{AP}^c)$ with $i \not = k$, there exists a $\tau' \in \mc{S}_i$
 such that $\tau' \not \in \mc{S}_k$.  So, $\Gamma$ can be obtained from the union of differences between each $\mc{S}_i$ ($i \not = k$) and $\mc{S}_k$.
}

\begin{example} \label{ex:abdex3}
Consider the instance $D$ with relations $R$ and $S$ as below, and the
\vspace{-9mm}
\begin{multicols}{2}
\noindent query $\Pi\!: \ \nit{ans} \leftarrow R(x, y), S(y)$,
which is true in $D$.
Assume all tuples are endogenous.

\vspace{3mm}
\begin{center} \begin{tabular}{l|c|c|} \hline
$R$  & A & B \\\hline
 & $a_1$ & $a_4$\\
& $a_2$ & $a_1$\\
& $a_3$ & $a_3$\\
 \hhline{~--}
\end{tabular} \hspace*{1cm}\begin{tabular}{l|c|c|}\hline
$S$  & B  \\\hline
 & $a_1$ \\
& $a_2$ \\
& $a_3$ \\ \hhline{~-}
\end{tabular}
\end{center}
\end{multicols}

In this case, $\mathcal{AP}^c= \langle \Pi, \emptyset, D, \nit{ans}\rangle$, which
has two (subset-minimal) abductive diagnoses:  $\Delta_1=\{S(a_1), R(a_2, a_1) \}$ and $\Delta_2=\{S(a_3), R(a_3, a_3)\}$. Then,
$\nit{Rel}(\mc{AP}^c)=\{S(a_3), $ $  R(a_3, a_3), $ $S(a_1), R(a_2, a_1)\}$.
It is easy to see that the relevant hypothesis are actual causes for $\nit{ans}$.
\boxtheorem
\end{example}

\subsection{Causal responsibility and abductive diagnosis}\label{sec:nessDeg}

In the previous section we showed that counterfactual and actual causes for Datalog query answers appear as necessary and relevant hypotheses in the associated Datalog abduction problem. The form
causal responsibility takes in Datalog abduction is less direct. \re{Actually, we first show that causal responsibility inspires an interesting concept for Datalog abduction, that of
{\em degree of necessity} of a hypothesis.}

 \ignore{That is $h$ is necessity if without which the  { DAP} obtained from removing $h$ from $\mc{AP}$ does not have any abductive diagnoses. This is, there is a counterfactual dependance between $h$ and the exitance of an abductive diagnosis for $\mc{AP}$}

\ignore{+++
 Recall from Definition \ref{def:DAP} that a hypothesis $h$ is necessity for a  { DAP} $\mc{AP}$, if it is contained in all diagnosis of $\mc{AP}$. We argue that this definition does not fully capture our intuition about necessary hypotheses for an abduction problem.
}
 \begin{example} \label{ex:nes} (ex. \ref{ex:abdex3} cont.) Consider now $D'=\{R(a_1, a_3), R(a_2, a_3), S(a_3)\}$, and $\mathcal{AP}^c= \langle \Pi, \emptyset, D', \nit{ans}\rangle$.
   \ $\mathcal{AP}^c$ has two abductive diagnosis:  $\Delta_1=\{S(a_3), R(a_1, a_3) \}$ and $\Delta_2=\{S(a_3), R(a_2, a_3)\}$.

    Here, $\nit{Ness}(\mathcal{AP}^c)=\{ S(a_3) \}$, i.e. only $S(a_3)$ is necessary for abductively explaining $\nit{ans}$. However, this is not capturing the fact that  $R(a_1, a_3)$ or $R(a_3, a_3)$ are also needed
    as a part of the explanation. \boxtheorem
 \end{example}
  This example suggests that necessary hypotheses might be better captured as sets of them rather than as individuals.

\begin{definition} \em  \label{def:nesshypset}
Given a  {DAP}, $\mathcal{AP}= \langle \Pi, E, \nit{Hyp}, \nit{Obs}\rangle$, \ignore{with
 \re{$\nit{Sol}(\mathcal{AP})\neq \emptyset$},} \ $N \subseteq \nit{Hyp}$ is a {\em necessary-hypothesis set}
 if: (a)  for $\mathcal{AP}_{\!\!-N} := \langle \Pi, E, \nit{Hyp} \smallsetminus N, \nit{Obs}\rangle$, $\nit{Sol}(\mathcal{AP}_\mn)=\emptyset$,
 and (b) $N$ is subset-minimal, i.e. no proper subset of $N$ has the previous property.
\boxtheorem
\end{definition}
It is easy to verify that a hypothesis
$h$ is necessary according to Definition \ref{def:DAP} iff
$\{h\}$ is a necessary-hypothesis set. 

If we apply Definition  \ref{def:nesshypset} to $\mathcal{AP}^c$ in Example \ref{ex:nes}, we obtain two necessary-hypothesis sets: $N_1=\{S(a_3)\}$ and $N_2=\{R(a_1, a_3), R(a_2, a_3)\}$.
 In this case, it makes sense to claim that  $S(a_3)$ is more necessary for explaining $\nit{ans}$ than the other two tuples, that need to be combined. \ignore{ This is because excluding this hypothesis makes the observation unexplainable within $\mathcal{AP}^c$. The other hypotheses do not have this property. This suggests that there is some dimension of necessity on which different hypotheses
occupy different positions.} Actually, we can think of ranking hypothesis according to the minimum cardinality of necessary-hypothesis sets where they are included.

\begin{definition} \em  \label{def:ness} Given a  {DAP}, $\mathcal{AP}= \langle \Pi, E, \nit{Hyp}, \nit{Obs}\rangle$,
 the {\em necessity-degree} of a hypothesis $h \in Hyp$ is
 $\eta_{_{\!\mc{AP}{}\!}}(h):= \frac{1}{|N|}$ where, $N$ is a minimum-cardinality necessary-hypothesis set with $h \in N$. If $h$
 does not belong to any necessary hypothesis set,  $\eta_{_{\!\mc{AP}{}\!}}(h):=0$.
\boxtheorem
\end{definition}
\begin{example} (ex. \ref{ex:nes} cont.) \ We have \
 $\eta_{_{\!\mc{AP}^c{}\!}}(S(a_3))=1$ and $\eta_{_{\!\mc{AP}^c{}\!}}(R(a_2, a_3))=\eta_{_{\!\mc{AP}^c{}\!}}(R(a_1, a_3))=\frac{1}{2}$.
Now, if we consider the original Datalog query in the causality setting, where $\Pi \cup D' \models \nit{ans}$,
then $S(a_3), R(a_2, a_3), R(a_1, a_3)$ are all actual causes, with responsibilities:
 $\rho_{_{\!\Pi\!}}^D(S(a_3))=1$,
 $\rho_{_{\!\Pi \!}}^D(R(a_2, a_3))$ $=\rho_{_{\!\Pi\!}}^D(R(a_1, a_3)) $ $=\frac{1}{2}$.
 This is not a
 coincidence. In fact the notion of causal responsibility is in correspondence with the notion of necessity degree in the Datalog abduction setting. \boxtheorem
 \end{example}
\begin{proposition} \label{pro:abdf&res}  \em Let $D=D^x \cup D^n$ be an instance and $\Pi$ be a Boolean Datalog query with $\Pi \cup D \models \nit{ans}$, and $\mc{AP}^c$ its associated CDAP. For $\tau \in D^n$, it holds:  $\eta_{_{\!\mc{AP}^{c}\!}}(\tau)=\rho_{_{\!\Pi\!}}^D(\tau)$.
\end{proposition}

\dproof{It is easy to verify that each  actual cause, together with a contingency set, forms a necessary hypothesis set for the corresponding causal Datalog abduction setting (and the other way around). Then, the two values are in correspondence.}

Notice that the notion of necessity-degree is interesting and
 applicable to general abduction from logical theories, that may not necessarily represent causal knowledge about a domain. In this case,
 the necessity-degree is not a causality-related notion, and merely  reflects the extent by which a hypothesis is necessary for making an observation explainable within an abductive theory.

\subsection{Abductive diagnosis from actual causes}

Now we show, conversely, that  QA-causality can capture Datalog abduction. In particular, we show that abductive diagnoses from Datalog programs are formed essentially by actual causes for the observation.
More precisely, consider a Datalog abduction problem $\mathcal{AP}= \langle \Pi, E, \nit{Hyp}, \nit{Obs}\rangle$, where $E$ is the
underlying extensional database, and
$\nit{Obs}$ is a conjunction of ground atoms.
For this we need to construct a
QA-causality setting. \ignore{ with $D := D^x \cup D^n$, $D^x:= E$, and $D^n:= \nit{Hyp}$, and the Datalog program
 $\re{\Pi^c} := \Pi \cup \{\nit{ans} \leftarrow \nit{Obs}\}$, with \nit{ans} a fresh propositional atom.
Then $\Pi^c$  can be seen as a monotone query on $D$. }

\begin{proposition} \label{pro:ac&rel}  \em Let $\mathcal{AP}= \langle \Pi, E, \nit{Hyp}, \nit{Obs}\rangle$ be a Datalog abduction problem, and $h \in \nit{Hyp}$. It holds that $h$ is a relevant hypothesis for $\mc{AP}$, i.e. $h \in \nit{Rel}(\mathcal{AP})$, iff $h$ is an actual cause for the associated Boolean Datalog query
$\Pi^c := \Pi \ \cup \ \{\nit{ans} \leftarrow \nit{Obs}\}$ being true in $D := D^x \cup D^n$ with
 $D^x:= E$, and $D^n:= \nit{Hyp}$. Here, \nit{ans} is a fresh propositional atom.
 \boxtheorem
\end{proposition}
The proof is similar to that of Proposition \ref{pro:abdf&cfcaus}, \red{where we start with a causality setting, producing an abductive setting. Instead, in this case we start from an abductive setting and produce a causal one.}

\begin{example} \label{ex:Datalogabduction} (ex.  \ref{ex:datalogabduction} cont.) \ For the given {DAP} $\mc{AP}$, we construct a QA-causality setting as follows. Consider the instance $D$ with relations
\nit{And}, \nit{Or}, $\nit{Faulty}$, \nit{One} and \nit{Zero}, as below, and the Boolean Datalog query $\Pi^c\!: \ \Pi \ \cup \ \{\nit{ans} \leftarrow \nit{Zero}({\sf d})\}$, where $\Pi$ is the
Datalog program in Example \ref{ex:datalogabduction}.

\vspace{-3mm}
\begin{multicols}{2}
\vspace*{-.2cm}
\begin{center} \begin{tabular}{l|c|c|c|c|} \hline
\nit{And} & $I_1$& $I_2$ & $O$ & $G$ \\\hline
& {\sf a} & {\sf b} & {\sf e}& {\sf and}\\
 \hhline{~----}
\end{tabular}

\vspace{2mm}
\hspace*{.5cm}\begin{tabular}{l|c|c|}\hline
$\nit{Zero}$  & I  \\\hline
& {\sf b} \\ \hhline{~-}
\end{tabular}

\vspace{2mm}
\begin{tabular}{l|c|c|c|c|} \hline
\nit{Or} & $I_1$& $I_2$ & $O$ & $G$ \\\hline
& {\sf e} & {\sf c} & {\sf d}& {\sf or}\\
 \hhline{~----}
 \end{tabular}

 \vspace{2mm}
 \hspace*{.8cm}\begin{tabular}{l|c|c|}\hline
$\nit{One}$  & I  \\\hline
& {\sf a} \\
& {\sf c} \\ \hhline{~-}
\end{tabular}~~~~
\begin{tabular}{l|c|c|}\hline
$\nit{Faulty}$  & $G$  \\\hline
& {\sf and} \\
& {\sf or} \\ \hhline{~-}
\end{tabular}
\end{center}

It clear that $\Pi^c \cup D \models \nit{ans}$.
 $D$ is partitioned into the set of endogenous tuples
$D^n:= \{\nit{Faulty({\sf and})},$ $\nit{Faulty({\sf or})}\}$ and the set of exogenous tuples $D^x := D \smallsetminus D^n$.

It is easy to verify that this result has only one actual cause, namely $\nit{Faulty({\sf or})}$ (with responsibility 1), confirming the correspondence with  Example \ref{ex:datalogabduction} as stated in Proposition \ref{pro:ac&rel}.
\boxtheorem
\end{multicols}
\end{example}

\subsection{Complexity of causality for Datalog queries} \label{sec:abdcomx}

Now we use the results obtained so far in this section to obtain new complexity results for Datalog  QA-causality. We first consider the problem of deciding if a tuple is a {\em counterfactual cause} for a query answer.

A counterfactual cause is a tuple that, when removed from the database, undermines the query-answer, without having to remove other tuples, as is the case for {\em actual causes}.
Actually, for each of the latter there may be an exponential number of contingency sets, i.e. of accompanying tuples \cite{icdt15}. Notice that a counterfactual cause is an actual cause with responsibility $1$.\ignore{
Given a set of counterfactual causes (assuming that they exist) for an unexpected/erronous query-answer, excluding any of them from the instance is a parsimonious recovery from the undesired situation. That is, the new instance does not imply the unexpected answer and obtained by removing only one tuple from the actual instance. If counterfactual causes exist and can be decided/computed in polynomial time, then the entire procedure to suppress the undesired answer would be tractable.}

\ignore{
\bblue{
This is in contrast with actual causes where, excluding a cause together with
one of its associated contingency sets undermines an answer. An actual cause may have exponential number of contingency sets \cite{icdt15}. Therefore, when an undesired answer only has actual causes, then the procedure to suppress the undesired answer may need to choose between exponential number of candidates to recover the system.}
}
\ignore{
Counterfactual causes are of particular importance because, the \bl{existence} of a counterfactual cause for an unexpected/erroneous answer \bblue{d} a polynomial-time recovery procedure to suppress the undesired answer. \bblue{More specifically, if an answer has opne o counterfactual causes
excluding any of the counterfactual causes for an answer from a database form, undermines the unexpected answer so its candidate recovery procure. This is in contract with actual causes where, each cause may have
This is because excluding a counterfactual cause for a query answer from a database, undermines the answer.}}

\ignore{
\comlb{I think the argument above is not the right one, in terms of polynomial-time procedure for the job. If you eliminate all the causes (which can be computed in polynomial time),
you get the same effect. With ramifications, of course, but you are not requesting anything other than not producing the answer anymore. Revise. The motivation as such is not good.}
}

\begin{definition} \em   \label{def:cfp}    For a  Boolean monotone query $\mc{Q}$,
the {\em counterfactual causality decision problem} ({CFDP}) is (deciding about membership of):

\vspace{1mm}
$\mc{CFDP}(\mc{Q}) := \{(D, \tau)~|~ \tau \in D^n \mbox{ and } \rho_{_{\!\mc{Q}\!}}^D(\tau)=1 \}.$ \boxtheorem
\end{definition}
The complexity of this problem can be obtained from the connection between counterfactual causation and the necessity of hypothesis in Datalog abduction via  Propositions \ref{pro:nessp} and \ref{pro:abdf&cfcaus}.

\begin{proposition}\label{pro:CFDP} \em
For Boolean Datalog queries $\Pi$, $\mc{CFDP}(\Pi)$ is in  {\nit{PTIME}} \ (in data).
\end{proposition}

\noindent
\dproof{Directly from Propositions \ref{pro:nessp} and \ref{pro:abdf&cfcaus}.}

Now we address the complexity of the actual causality problem for Datalog queries. The following result is obtained from  Propositions \ref{pro:relp} and  \ref{pro:ac&rel}.

\begin{proposition}\label{pro:cp} \em
For Boolean Datalog queries $\Pi$, $\mc{CDP}(\Pi)$ is {\em  {NP}}-complete (in data).
\end{proposition}

\dproof{To show the membership of {\em NP}, consider an instance $D=D^n \cup D^x$ and a tuple $\tau \in D^n$. To check if
$(D, \tau) \in  \mc{CDP}(\Pi)$ (equivalently $\tau \in \nit{Causes}(D, \Pi))$, non-deterministically guess
 a subset $\Gamma \subseteq D^n$, return {\it yes} if $\tau$ is a counterfactual cause for $\mc{Q}(\bar{a})$ in  $D  \smallsetminus \Gamma$, and \nit{no} otherwise.
By Proposition \ref{pro:CFDP} this can be done in polynomial time.

The {\em NP}-hardness is obtained by a reduction from the relevance problem for Datalog abduction to causality problem, as given in Proposition \ref{pro:ac&rel}.}

\vspace{2mm}
This result should be contrasted with the tractability of the same problem for {UCQ}s \cite{icdt15}. \re{In the case of Datalog, the {\em NP}-hardness requires a recursive query. This can be seen from
the proof of Proposition \ref{pro:cp}, which  appeals in the end
to the {\em NP}-hardness in Proposition \ref{pro:relp}, whose proof uses a recursive query (program) (cf. the query given by (\ref{eq:red1})-(\ref{eq:red2}) in the Appendix).}

We now introduce a fixed-parameter tractable case of the actual causality problem. Actually, we consider the ``combined" version of the decision problem
in Definition \ref{def:cp}, where both the Datalog query and the instance are part of the input.
For this, we take advantage of the tractable case of Datalog abduction presented in Section \ref{sec:backAbd}.
The following is an immediate consequence of Theorem \ref{the:trac} and  Proposition \ref{pro:abdf&cfcaus}.


\begin{proposition}\label{cor:traccaus} \em
For a guarded Boolean Datalog query $\Pi$, an instance $D = D^x \cup D^n$, with $D^x$ of bounded tree-width, and $\tau \in D^n$, deciding if $\tau \in \nit{Causes}(D,\Pi)$ is fixed-parameter tractable (in combined complexity), and
the parameter is the tree-width bound.
\boxtheorem
\end{proposition}
Finally, we establish the complexity of the responsibility problem for Datalog queries.

\begin{proposition}\label{pro:rp} \em
For Boolean Datalog queries $\Pi$, $\mc{RDP}(\Pi)$ is {\em  {NP}}-complete.
 \boxtheorem
\end{proposition}

\dproof{ To show  membership of {\em NP}, consider an instance $D=D^n \cup D^x$, a tuple $\tau \in D^n$, and a responsibility bound $v$. To
check if $\rho_{_{\!\Pi\!}}^D(\tau)>v$, non-deterministically guess a set $\Gamma \subseteq D^n$ and check if $\Gamma$ is a contingency set and $\Gamma < \frac{1}{v}$.
The verification can be done in polynomial time. Hardness is obtain from the
{\em NP}-completeness of  {RDP} for conjunctive queries established in \cite{icdt15}.
}

\section{Causality and View-Updates} \label{sec:delp&cause}

There is a close relationship between QA-causality   and the view-update problem in the form of delete-propagation. It was first suggested in  \cite{Kimelfeld12a, Kimelfeld12b}, and here we investigate
it more deeply.
We start by formalizing some  computational problems related to the general delete-propagation problem that are interesting from the perspective of QA-causality.

\subsection{Background on delete-propagation}\label{sec:del-pro}

Given a monotone query $\mc{Q}$, we can think of it as defining a view, $\mc{V}$, with virtual contents $\mc{Q}(D)$. If $\bar{a} \in \mc{Q}(D)$, which may not be intended,
we  may try to delete some tuples from $D$, so that  $\bar{a}$  disappears from $\mc{Q}(D)$. This is a particular case of database updates through views
\cite{Abiteboul95}, and may appear in different and natural formulations. The next example shows one of them.

\begin{example} \label{ex:dpexample2} \  
Consider relational predicates \
$\nit{GroupUser(User, Group})$ \ and $\nit{GroupFile(File,Group})$, with extensions as in instance $D$ below. They represent users' memberships of groups, and access permissions for groups to files,
respectively.\footnote{ \ This example, originally presented in \cite{CuiWW01} and later used in \cite{BunemanKT02, Kimelfeld12a, Kimelfeld12b}, is borrowed from the area of view-updates. We use it here to point to the similarities
 between the seemingly different problems of view-updates and causality. \ignore{We elaborate on this in \re{Section \ref{sec:delp&cause}}.}}

\vspace{0.3cm}
\begin{center}
{\small
\begin{tabular}{l|c|c|} \hline
\nit{GroupUser} & \ \nit{User} \ & \nit{Group} \\\hline
 & {\sf Joe}  & $g_1$\\
 & {\sf Joe}  & $g_2$\\
& {\sf John} & $g_1$\\
& \sf{Tom} & $g_2$\\
& \sf{Tom} & $g_3$\\
& {\sf John} & $g_3$\\
 \hhline{~--} \end{tabular} ~~~~~~~\begin{tabular}{l|c|c|} \hline
\nit{GroupFiles}  & \nit{File} & \nit{Group} \\\hline
 & $f_1$ & $g_1$ \\
  & $f_1$ & $g_3$ \\
& $f_2$ & $g_2$\\
& $f_3$& $g_3$\\
 \hhline{~--}
\end{tabular}
}
\end{center}

\vspace{2mm}

It is expected that a user $u$ can access  file
$f$ if $u$ belongs to a group that can access $f$, i.e. there
is some group $g$ such that $\nit{GroupUser}(u, g)$  and $\nit{GroupFile}(f,g)$ hold.
Accordingly, we can define a view  that collects users with the files they can access, as defined by the following query:
\begin{equation}\label{eq:access}
\nit{Access(User, File}) \ \leftarrow \ \nit{GroupUser(User, Group), GroupFile(File, Group}).
\end{equation}

Query $\nit{Access}$ in (\ref{eq:access}) has the following answers, providing a view extension:

\vspace{2mm}
\begin{multicols}{2}
{\small
\begin{tabular}{l|c|c|} \hline
$\nit{Access}(D)$ & \nit{User} & \nit{File} \\\hline
 & {\sf Joe}  & $f_1$ \\
  & {\sf Joe}  & $f_2$\\
& \sf{Tom} & $f_1$\\
& \sf{Tom} & $f_2$\\
& \sf{Tom} & $f_3$\\
& {\sf John} & $f_1$\\
& {\sf John} & $f_3$\\
 \hhline{~--}
 \end{tabular}
}

\phantom{ooo}

\noindent In a particular version of  the delete-propagation problem, the objective may be to delete a {\em  minimum number} of tuples from the instance, so that an authorized access (unexpected answer to the query) is deleted from the query answers, while all other authorized accesses (other answers to the query) remain intact. \boxtheorem
\end{multicols}
\end{example}
In the following, we consider several variations of this problem, both in their functional and decision versions.

\begin{definition} \em \label{def:SubSDP}  Let $D$ be a database instance, and $\mc{Q}(\bar{x})$ a monotone query.
\begin{enumerate}[(a)]
\item For $\bar{a} \in \mc{Q}(D)$, the {\em minimal-source-side-effect deletion-problem} is about computing a subset-minimal $\Lambda \subseteq D$, such that $\bar{a} \ \notin \mc{Q}(D \smallsetminus \Lambda)$.

\item The {\em minimal-source-side-effect decision problem}  is (deciding about the membership \ of):

\vspace{1mm}
 $\mc{MSSEP}^{s\!}(\mc{Q})= \{(D, D', \bar{a}) ~|~ \bar{a}\in \mc{Q}(D), \ D' \subseteq D, \bar{a} \not \in \mc{Q}(D'), \mbox{ and} $\\
 \hspace*{8cm}$ D' \mbox{ is subset-maximal} \}$.\\
 {\small (The superscript $s$ stands for subset-minimal.)}

\item For $\bar{a} \in \mc{Q}(D)$, the {\em minimum-source-side-effect deletion-problem} is about computing a minimum-cardinality $\Lambda \subseteq D$, such that $\bar{a} \notin \mc{Q}(D \smallsetminus \Lambda)$.

\item The {\em minimum-source-side-effect} {\em decision problem} is (deciding about the membership of):

\vspace{1mm}
$\mc{MSSEP}^{c\!}(\mc{Q})= \{(D, D', \bar{a}) ~|~ \bar{a}\in \mc{Q}(D), D' \subseteq D, \ \bar{a} \notin \mc{Q}(D'),  \mbox{ and }$\\
 \hspace*{6.5cm} $ D' \mbox{ has maximum cardinality} \}$.\\
{\small (Here, $c$ stands for  cardinality.)}
\boxtheorem
\end{enumerate}
\end{definition}


\begin{definition} \em \label{def:FreeVDP} \cite{BunemanKT02} \ Let $D$ be a database instance $D$, and $\mc{Q}(\bar{x})$ a  monotone query.
\begin{enumerate}[(a)]
\item For $\bar{a} \in \mc{Q}(D)$, the {\em view-side-effect-free deletion-problem} is about computing a  \ignore{minimum-cardinality} $\Lambda \subseteq D$, such that $\mc{Q}(D) \smallsetminus \{\bar{a}\} = \mc{Q}(D
\smallsetminus \Lambda)$.

\item The {\em
view-side-effect-free decision problem} is (deciding
about the membership of):

\vspace{1mm}
$\mc{VSEFP}(\mc{Q})= \{(D, \bar{a}) ~|~  \bar{a} \in
\mc{Q}(D), \ \mbox{and exists } D' \subseteq D \mbox{ with }$ \\ \hspace*{7.5cm}$ \mc{Q}(D) \smallsetminus
\{\bar{a}\} = \mc{Q}(D')\} $. \boxtheorem
\end{enumerate}
\end{definition}


\re{The decision problem in Definition \ref{def:FreeVDP}(b) is \nit{NP}-complete for conjunctive queries \cite[theorem 2.1]{BunemanKT02}. }
\re{Notice that, in contrast to those in (a) and (c) in Definition \ref{def:SubSDP}}, this decision problem does not involve a candidate $D'$, and only asks about its existence. This is because
candidates always exist for Definition \ref{def:SubSDP}, whereas for the {\em view-side-effect-free deletion-problem} there may be no sub-instance that produces {\em exactly} the intended deletion
from the view. \re{As usual, there are   {\em functional} problems associated to $\mc{VSEFP}$, about
computing a maximal/maximum $D'$ that produces the intended side-effect free deletion; and also the two corresponding decision problems about deciding concrete candidates $D'$.}


\begin{example} \label{ex:dpp} (ex. \ref{ex:cfex1} cont.) \ Consider the instance $D$ and the  conjunctive query $\mc{Q}$ in (\ref{eq:query}). Assume that {\sf XML} is not among {\sf John}'s research interests, so that  tuple $\langle {\sf John},{\sf XML} \rangle$ in the
view $\mc{Q}(D)$ is unintended. We want to find tuples in $D$ whose removal
leads to the deletion of this view tuple. There are multiple ways  to achieve this goal.

Notice that  the tuples in $D$ related to answer $\langle {\sf John},{\sf XML} \rangle$ through the query are  \nit{Author({\sf John}, {\sf TKDE})}, \nit{Journal}({\sf TODS}, {\sf XML}, \mbox{32}),
 \nit{Author}({\sf John}, {\sf TODS}) and  \nit{Journal}({\sf TKDE}, {\sf XML},\mbox{30}). They are all candidates for removal. However,
 the decision problems described above impose
different conditions on what are admissible deletions.

\noindent (a) Source-side effect:  The objective is to find minimal/minimum sets of tuples whose removal leads to the deletion of $\langle \nit{\sf John},{\sf {\sf XML}} \rangle$. One solution is removing  $S_1$=\{\nit{Author({\sf John}, {\sf TODS}), } \nit{Author({\sf John}, {\sf TKDE})}\}  from the \nit{Author} table. The other solution is  removing $S_2$=\{\nit{Journal}({\sf TODS}, {\sf XML},\mbox{30}), \nit{Journal}({\sf TKDE}, {\sf XML}, \mbox{  30})\} from the \nit{Journal} table.

Furthermore,
the removal of either $S_3=\{\nit{Author}({\sf John}, {\sf TKDE}),$ \linebreak $ \nit{Journal}({\sf TODS},$ $ {\sf XML}, \mbox{32})\}$ or  $S_4=\{\nit{Author}({\sf John}, {\sf TODS}), \nit{Journal}({\sf TKDE}, {\sf XML},$ $\mbox{30})\} $ eliminates the intended view tuple.
Thus, $S_1$, $S_2$, $S_3$ and $S_4$ are solutions to both the minimum- and minimal-source side-effect deletion-problems.

\noindent (b) View-side effect: Removing any of the sets $S_1$, $S_2$, $S_3$ or $S_4$, leads to the deletion
of $\langle {\sf John},{\sf XML} \rangle$. However, we now want those sets whose elimination produce
no side-effects on the view. That is,  their deletion triggers the
deletion of $\langle {\sf John},{\sf XML} \rangle$ from the view, but not of any other tuple in it.

None of the sets $S_1$, $S_2$, $S_3$ and $S_4$  is side-effect free. For example, the deletion of $S_1$
also results in the deletion of $\langle {\sf John}, {\sf CUBE} \rangle$ from the view. \boxtheorem
\end{example}

\begin{example} \label{ex:dpexample3} (ex. \ref{ex:dpexample2} cont.) \
It is easy to verify that there is no solution to the view-side-effect-free deletion-problem for answer $\langle {\sf Tom}, f_3 \rangle$ (in the view extension
$\nit{Access}$). To eliminate this entry
from the view, either $\nit{GroupUser}({\sf Tom}, g_3)$ or $\nit{GroupFiles}(f_3,g_3)$ must be deleted from $D$. Removing the former
results in the additional deletion of $\langle {\sf Tom}, f_1 \rangle$ from the view; and eliminating the latter, results in the additional deletion of  $\langle {\sf John}, f_3 \rangle$.

However, for the answer $\langle {\sf Joe}, f_1 \rangle$,  there is a solution to the  view-side-effect-free deletion-problem, by removing $\nit{GroupUser}({\sf Joe}, g_1)$ from $D$. This deletion does not have
unintended side-effects on the view contents. \boxtheorem
\end{example}

\subsection{QA-causality and  delete-propagation} \label{sec:causeanddp}

In this section we first establish mutual reductions between the   delete-propagation problems and QA-causality.


\subsubsection{Delete propagation from QA-causality. \ }

{\bf \em In this section, unless otherwise stated, all the database tuples are assumed to be endogenous.\footnote{ \ The reason is that in this section we want to characterize view-deletions
 in term of causality, but for the former problem we did not partition the database tuples.}}

 Consider a relational instance $D$, a view $ \mc{V}$ defined by a monotone query $\mc{Q}$. Then, the virtual view extension, $\mc{V}(D)$, is $\mc{Q}(D)$. 

For a tuple $\bar{a} \in \bblue{\mc{Q}}(D)$, the delete-propagation problem, in its most general form, is about deleting a set of tuples from $D$, and so obtaining a subinstance $D'$ of $D$, such that
$\bar{a} \notin \mc{Q}(D')$. It is natural to expect that the deletion of $\bar{a}$ from $\bblue{\mc{Q}}(D)$
can be achieved through deletions from $D$ of actual causes for $\bar{a}$ (to be in the view extension).
However, to obtain solutions to the different variants of this problem introduced in Section \ref{sec:del-pro}, different combinations  of actual causes must be considered.

First, we show that an actual cause for $\bar{a}$ \ignore{to be in $\mc{V}(D)$} forms, with any of its contingency sets, a solution to the minimal-source-side-effect deletion-problem associated to $\bar{a}$ (cf. Definition \ref{def:SubSDP}).


\begin{proposition} \label{pro:causeSubSDP} \em
 For an instance $D$, a subinstance $D' \subseteq D$, a view \ignore{$ \mc{V}$} defined by a monotone query $\mc{Q}(\bar{x})$, and  $\bar{a}
 \in \mc{Q}(D)$, \  $(D, D', \bar{a}) \in \mc{MSSEP}^{s\!}(\mc{Q})$ \ iff \ there is a $\tau
 \in D \smallsetminus D'$, such that  $ \tau \in\nit{Causes}(D, \mc{Q}(\bar{a}))$ and $ D \smallsetminus (D' \cup  \{\tau\}) \in \nit{Cont}^s(D, \mc{Q}(\bar{a}), \tau)$.
\end{proposition}

\dproof{Suppose first that $(D, D', \bar{a}) \in \mc{MSSEP}^{s\!}(\mc{Q})$. Then, according to Definition \ref{def:SubSDP},
$\bar{a} \not \in \mc{Q}(D')$. Let $\Lambda= D \smallsetminus D'$. For an
arbitrary element $\tau \in \Lambda$ (clearly, $\Lambda \not =\emptyset$), let $\Gamma:= \Lambda \smallsetminus \{\tau\}$. Due to the subset-maximality of
$D'$ (then, subset-minimality of $\Lambda$), we obtain: $D  \smallsetminus (\Gamma \cup \{\tau\}) \not \models \mc{Q}(\bar{a})$, but $D  \smallsetminus \Gamma  \models \mc{Q}(\bar{a})$. Therefore, $\tau$ is an actual cause for $\bar{a}$.

For the other direction,  suppose $ \tau \in\nit{Causes}(D, \mc{Q}(\bar{a}))$ and $ D \smallsetminus (D' \cup  \{\tau\}) \in \nit{Cont}^s(D, \mc{Q}(\bar{a}), \tau)$. Let $\Gamma :=D \smallsetminus (D' \cup  \{\tau\})$. From the definition
of an actual cause, we obtain that  $\bar{a} \ \notin \mc{Q}(D \smallsetminus (\Gamma \cup \{\tau\})$. So, $\bar{a} \ \notin \mc{Q}(D')$ (notice that $D'=D \smallsetminus (\Gamma \cup \{\tau\}$). Since $\Gamma$ is a subset-minimal contingency set for $\tau$,
 $D'$ is a subset-maximal subinstance that enjoys the mentioned property. So, $(D, D', \bar{a})
 \in \mc{MSSEP}^{s\!}(\mc{Q})$.
}

\begin{corollary} \em For a view defined by a monotone query, deciding if a set of source deletions producing the  deletion from the view is subset minimal
is in polynomial time in data.
\end{corollary}
\dproof{This follows from the connection between QA-causality and delete-propagation established in Proposition \ref{pro:causeSubSDP}, and the fact that deciding a cause for a monotone query and deciding the subset minimality  of an associated
contingency set candidate are both in polynomial time in data \cite{Meliou2010a,tocs15}.}

\ignore{
\re{Through the just established connection, and the fact that deciding a cause for a monotone query and deciding the subset minimality  of an associated
contingency set candidate are both in polynomial time in data \cite{Meliou2010a,tocs15}, we obtain that deciding if a set of source deletions producing the  deletion from the view is subset minimal
is in polynomial time in data.} }

\ignore{\comlb{I added this paragraph above. I did not find anything explicit in Buneman's paper about checking candidates. But, juts in case, I am not putting it in a formal
result environment. Please check on your side too.} }

We show
next that, in order to minimize {\em the number} of side-effects on the source (the problem in Definition \ref{def:SubSDP}(c)),
it is good enough to pick a most responsible cause for $\bar{a}$ with any of its minimum-cardinality contingency sets.


\begin{proposition} \label{pro:causeminSDP} \em
  For an instance $D$, a subinstance $D' \subseteq D$, a view $ \mc{V}$ defined by a monotone query $\mc{Q}$, and  $\bar{a} \in \mc{Q}(D)$, \  $(D, D', \bar{a}) \in
\mc{MSSEP}^{c\!}(\mc{Q})$ \ iff \ there is  a $\tau \in D \smallsetminus D'$, such
that  $\tau \in  \mc{MRC}(D, \mc{Q}(\bar{a}))$, $ \Gamma:= D \smallsetminus (D'
\cup \{\tau\}) \in \nit{Cont}^s(D, \mc{Q}(\bar{a}), \tau)$, and there is no $\Gamma'
\in \nit{Cont}^s(D, \mc{Q}(\bar{a}), \tau)$ with $|\Gamma'|< |\Gamma|$.
\end{proposition}

\dproof{Similar to the proof of Proposition \ref{pro:causeSubSDP}.}

\re{In relation to the problems involved in this proposition,  the decision problems associated to computing a minimum-side-effect source deletion and computing the responsibility of a cause,
both for monotone queries,
have been independently established as {\em NP}-complete in data, in  \cite{BunemanKT02} and \cite{tocs15}, resp.}



\begin{example}\label{ex:Vpex1} (ex. \ref{ex:dpp} cont.)
We obtained the followings solutions to the minimum- (and also minimal-) source-side-effect deletion-problem for the view tuple $\langle\nit{{\sf John},{\sf XML}}\rangle$:
\begin{eqnarray*}
S_1&=&\{\nit{Author({\sf John}, {\sf TODS}), Author({\sf John}, {\sf TKDE})}\},\\
S_2&=&\{\nit{Journal}({\sf TODS}, {\sf XML},\mbox{30}), Journal({\sf TKDE}, {\sf XML},\mbox{30})\},\\
S_3&=&\{\nit{Author}({\sf John}, {\sf TKDE}), Journal({\sf TODS}, {\sf XML},\mbox{32})\}, \\
S_4&=&\{\nit{Author}({\sf John}, {\sf TODS}), Journal({\sf TKDE}, {\sf XML},\mbox{30})\}.
\end{eqnarray*}
On the other side, in Example \ref{ex:cfex1}, we  showed that the tuples \nit{Author({\sf John},{\sf TODS})}, \red{\nit{Journal}(\mbox{\sf TKDE},\mbox{\sf XML},\mbox{30})}, \nit{Author(\mbox{\sf John},\mbox{\sf TKDE})}, and \nit{Journal}({\sf TODS}, {\sf XML},\mbox{32}) are actual causes  for the answer $\langle \nit{\sf John},{\sf XML} \rangle$ (to the view query). In particular, for the cause \nit{Author({\sf John},{\sf TODS})} we obtained two contingency sets:  $\Gamma_1 = \{\nit{Author(\mbox{\sf John},\mbox{\sf TKDE})}\}$
 and $\Gamma_2$=\{\red{\nit{Journal}(\mbox{\sf TKDE},\mbox{\sf XML},\mbox{30})\}}.

It is easy to verify that each actual cause for answer $\langle\nit{{\sf John},{\sf XML}}\rangle$, together with any of its subset-minimal (and minimum-cardinality) contingency sets, forms a solution to
the minimal- (and minimum-) source-side-effect deletion-problem for $\langle\nit{{\sf John},{\sf XML}}\rangle$. For illustration, $\{\nit{Author(\mbox{\sf John}, \mbox{\sf TODS})}\} \cup \Gamma_1$ coincides with  $S_1$,
and  $\{\nit{Author(\mbox{\sf John}, \mbox{\sf TODS})} \} \cup \Gamma_1$ coincides with $S_4$. Thus, both of them are solutions to
 minimal- (and minimum-) source-side-effect deletion-problem for the view tuple $\langle\nit{{\sf John},{\sf XML}}\rangle$.  This confirms Propositions \ref{pro:causeSubSDP} and
\ref{pro:causeminSDP}. \boxtheorem
\end{example}

\re{Now we consider a variant of the functional problem in Definition \ref{def:SubSDP}(c), about computing {\em the minimum number} of source deletions.}
The next result is obtained from  the $\nit{FP}^\nit{ {NP}(log(n))}$-completeness of computing the highest responsibility associated to a query answer (i.e. the responsibility of the most
responsible causes for the answer)  \cite[prop. 42]{icdt15}.

\begin{proposition} \label{pro:asep} \em
Computing the size of a solution to \bblue{a} minimum-source-side-effect deletion-problem
 is $\nit{FP}^\nit{ {NP}(log(n))}$-hard. \boxtheorem
\end{proposition}

\dproof{By reduction from computing responsibility of a most responsible
cause (cf. Definition \ref{def:mracp}) via the characterization in Proposition \ref{pro:causeminSDP}.}

\subsubsection{QA-causality from  delete-propagation.}

\re{{\em In this subsection we  assume that all tuples are endogenous since the endogenous vs. exogenous classification has not been considered on the view update side} (but cf. Section \ref{sec:viewsEnd}).}

Consider a relational instance $D$, and a monotone query $\mc{Q}$ with $\bar{a} \in \mc{Q}(D)$. We will show that actual causes and most responsible causes for  $\bar{a}$ can be obtained from different variants of the delete-propagation problem associated with  $\bar{a}$.

First, we show that actual causes for a query answer can be obtained from the solutions to a corresponding minimal-source-side-effect deletion-problem.

\begin{proposition} \label{pro:causefromview} \em
For an instance $D$ and a monotone query $\mc{Q}(\bar{x})$ with $\bar{a} \in \mc{Q}(D)$, $\tau \in D$ is an actual cause for $\bar{a}$ \ iff \ there is a $D' \subseteq D$ with $\tau\in  (D\smallsetminus D')$
and $(D, D', \bar{a}) \in \mc{MSSEP}^{s\!}(\mc{Q})$.
\end{proposition}

\dproof{ Suppose $\tau \in D$ is an actual cause for $\bar{a}$ with a subset-minimal contingency set $\Gamma \subseteq D$. Let $\Lambda= \Gamma \cup \{\tau\}$ and $D'=D \smallsetminus \Lambda$ .  It is clear that
$\bar{a} \not \in \mc{Q}(D')$. Then, due to the subset-minimality of $\Lambda$, we obtain that $(D, D', \bar{a}) \in \mc{MSSEP}^{s\!}(\mc{Q})$.  A
similar argument applies to the other direction.
}

Similarly, most-responsible causes for a query answer can be obtained from solutions to a corresponding minimum-source-side-effect deletion-problem.
\begin{proposition} \label{pro:mostrescausefromview} \em
 For an instance $D$ and a monotone query $\mc{Q}(\bar{x})$ with $\bar{a} \in \mc{Q}(D)$,  $\tau \in D$   is a most responsible actual cause for $\bar{a}$ \ iff \ there is a $D' \subseteq D$ with $t \in (D\smallsetminus D')$  and  $(D, D', \bar{a}) \in \mc{MSSEP}^{c\!}(\mc{Q})$.
\end{proposition}

\dproof{ Similar to the proof of Proposition \ref{pro:causefromview}.}


\begin{example} \label{ex:causefromview} (ex. \ref{ex:cfex1} and \ref{ex:Vpex1} cont.) \
Assume all tuples are endogenous. \ We obtained $S_1$, $S_2$, $S_3$ and $S_4$ as solutions to the minimal- (and minimum-) source-side-effect deletion-problems for the view-element $\langle\nit{\sf John},{\sf XML}\rangle$.
Let $S$ be their union, i.e. $S=\{\nit{Author(\mbox{\sf John}, \mbox{\sf TODS})},  \red{\nit{Journal}(\mbox{\sf TKDE},\mbox{\sf XML},\mbox{30})}, \nit{Author}(\mbox{\sf John},$ $\mbox{\sf TKDE}),  \nit{Journal}({\sf TODS}, {\sf XML},\mbox{32})\}$.

We can see that $S$ contains actual causes for $\langle\nit{\sf John},{\sf XML}\rangle$. In this case, actual causes are also most responsible causes. This coincides with the
results obtained in Example \ref{ex:cfex1}, and confirms Propositions \ref{pro:causefromview} and \ref{pro:mostrescausefromview}.
\boxtheorem
\end{example}

Consider a view defined by a query $\mc{Q}$ as in Proposition \ref{pro:mostrescausefromview}. Deciding if a candidate contingency set (for an actual cause $\tau$) has minimum cardinality (giving to  $\tau$ its responsibility value) is {\em the complement} of checking if a set of tuples
is a maximum-cardinality repair (i.e  a {\em cardinality-based repair} \cite{2011Bertossi}) of the given instance  with respect to the denial constraint that has
$\mc{Q}$ as violation view (instantiated on $\tau$). The latter problem is in {\em coNP}-hard in data \cite{icdt07,kolaitis}. Thus, we obtain that checking minimum-cardinality contingency sets is  {\em NP}-hard in data.
Appealing to Proposition \ref{pro:mostrescausefromview}, we can reobtain via repairs and causality the result in \cite{BunemanKT02} about the $\nit{NP}$-completeness of $\mc{MSSEP}^{c\!}(\mc{Q})$. We illustrate the connection with an example.

\begin{example}  Consider the  instance $D$ as below, and the view $V$ defined by

\vspace{-4mm}
\begin{multicols}{2}
\noindent the query \ $V(y) \ \leftarrow \ R(x, y), S(y)$.

A view element (and query answer) is:  $\langle a_1 \rangle$.

Now, the denial constraint that has this (instantiated) view  as violation view is
\ $\kappa\!: \ \neg V(a_1)$, equivalently,

\vspace{5mm}
\phantom{oo} \vspace{-3mm}
\begin{center} \begin{tabular}{l|c|c|} \hline
$R$  & A & B \\\hline
 & $a_1$ & $a_4$\\
& $a_2$ & $a_1$\\
& $a_3$ & $a_1$\\
 \hhline{~--}
\end{tabular} \hspace*{1cm}\begin{tabular}{l|c|c|}\hline
$S$  & B  \\\hline
 & $a_1$ \\
& $a_2$ \\
& $a_3$ \\ \hhline{~-}
\end{tabular}
\end{center}
\end{multicols}
\vspace{-2mm} \noindent $\kappa\!: \ \neg \exists x (R(x, a_1) \wedge S(a_1))$. Instance $D$ is inconsistent with respect to $\kappa$, and has to be repaired by keeping a consistent subset of $D$ of maximum cardinality. The only cardinality-repair is: $D\smallsetminus \{S(a_1)\}$. The complement of this repair,
 $\Gamma = \{S(a_1)\}$, will be the minimum-cardinality contingency set for any cause in $D$ for the query answer, i.e. for $R(a_2,a_1)$ and $R(a_3,a_1)$, but not for the cause $S(a_1)$, which is
a counterfactual cause. Cf. \cite{tocs15} for more details on the relationship between repairs and causes with their contingency sets. \boxtheorem
\end{example}

\ignore{
\re{Through the just established connection, and the fact that deciding the  minimality in cardinality of a
contingency set candidate is {\em coNP}-hard in data \cite{tocs15}, we obtain that deciding if a set of source deletions producing the  deletion from the view is  minimal
in cardinality is {\em coNP}-hard in data.}
\comlb{Idem. Actually, in Buneman's paper there is not much about cardinality. And we get coNP from the cardinality-repair connection. Why didn't we have any explicit coNP-hardness results in ICDT15? Please, check.}
}

\section{View-Conditioned Causality} \label{sec:vcc} 

\subsection{VC-causality and its decision problems}

 QA-causality is defined for a fixed query $\mc{Q}$ and a fixed answer $\bar{a}$.  However, in practice one often has multiple queries and/or multiple answers. For a query with several answers one might be interested in causes for a fixed answer, on the condition that  the other query answers are correct. This
 form of {\em conditioned  causality} was suggested in \cite{Meliou2010b}; and formalized in \cite{Meliou2011}, in a more general, non-relational setting, to give an account of the effect of a tuple on multiple outputs (views). Here we adapt this notion of {\em view-conditioned causality} to the case of a single query, with possibly several answers. We illustrate first the  notion with a couple of examples.

\begin{example}\label{ex:motvcex1} (ex. \ref{ex:cfex1} cont.)\ignore{(ex. \ref{ex:dpp} cont.)} \ Consider again the answer $\langle \nit{\sf John},{\sf XML}\rangle$ to $\mc{Q}$. Suppose this answer is unexpended and likely to be wrong, while all other answers to $\mc{Q}$ are known to be correct. In this case, it makes sense that for the causality status of $\langle \nit{\sf John},{\sf XML}\rangle$ only those contingency sets  whose removal does not affect the correct answers to
the query are admissible. In other words, the hypothetical states of the database $D$ that do not provide the correct answers are not considered.
 \boxtheorem
\end{example}

\begin{example}\label{ex:motivcex2} (ex. \ref{ex:dpexample2} cont.)
Consider the  query in (\ref{eq:access}) as defining a view $\nit{Access}$, collecting users and the files they can access.

Suppose we observe that a particular file is accessible by an unauthorized user (an unexpected answer to the query), while all other users' accesses  are
known to be authorized (i.e. the other answers to the query are deemed to be  correct). We want to find out the causes for this unexpected observation. For this task, contingency sets whose removal do not return
the correct answers anymore should not be considered.
 \boxtheorem
\end{example}

 More generally, consider  a query $\mc{Q}$ with $\mc{Q}(D)=\{\bar{a}_1, \ldots, \bar{a}_n\}$. Fix an answer, say $\bar{a}_1 \in \mc{Q}(D)$, while the other answers  will be used as a condition on
 $\bar{a}_1$'s causality. Intuitively, $\bar{a}_1$ is somehow unexpected, we look for causes, but considering the other answers as ``correct". This has the effect of
 reducing the spectrum of contingency sets, by keeping $\mc{Q}(D)$'s extension fixed (the fixed {\em view extension}), except for $\bar{a}_1$ \ \citep{Meliou2011}.

 \begin{definition} \label{def:vccause} \em Given an instance $D$ and a monotone query $\mc{Q}$, consider $\bar{a} \in \mc{Q}(D)$, and \re{$V := \mc{Q}(D) \smallsetminus \{\bar{a}\}$}:
 \begin{itemize}
 \item[(a)] Tuple $\tau \in D^n$ is a {\em view-conditioned counterfactual cause} ({vcc-}cause) {\em for}  $\bar{a}$ \re{\em in $D$ relative to $V$} if  $\bar{a} \notin \mc{Q}(D \smallsetminus \{\tau\})$ but $\mc{Q}(D \smallsetminus\{\tau\}) = V$.

 \item[(b)] Tuple $\tau \in D^n$  is a {\em view-conditioned actual cause} ({vc-}cause) {\em for}  $\bar{a}$ {\em in $D$ relative to $V$} if there  exists a contingency set, $\Gamma \subseteq D^n$,
 such that $\tau$ is a   {vcc}-cause for $\bar{a}$ in $D \smallsetminus \Gamma$ relative to $V$.

 \item[(c)]
 $\nit{vc\mbox{-}\!Causes}(D, \mc{Q}(\bar{a}))$ denotes the set of all  {vc-}causes for $\bar{a}$.
 \item[(d)]
  The {\em {vc-}causal responsibility} of a tuple $\tau$ for answer $\bar{a}$ is  $\nit{vc\mbox{-}}\rho_{_{\!\mc{Q}(\bar{a})\!}}^D(\tau) := \frac{1}{1 + |\Gamma|}$, where $|\Gamma|$ is the
size of the smallest contingency set that makes $\tau$ a vc-cause for  $\bar{a}$. \boxtheorem
\end{itemize}
\end{definition}
Notice that the implicit conditions on vc-causality in Definition \ref{def:vccause}(b) are: $\bar{a} \in \mc{Q}(D \smallsetminus \Gamma)$, \ $\bar{a} \notin (\mc{D} \smallsetminus (\Gamma \cup \{\tau\}))$, \ and  $\mc{Q}(D \smallsetminus (\Gamma \cup \{\tau\})) = V$. In the following, we will omit saying ``relative to $V$" since the fixed contents can be understood from the context.

Clearly, $ \nit{vc\mbox{-}\!Causes}(D, Q(\bar{a})) \subseteq \nit{Causes}(D, Q(\bar{a}))$,  but not necessarily the other way around.
Furthermore,  the causal responsibility and the {vc-}causal responsibility
 of a  tuple as a cause, resp. vc-cause, for a same query answer may take different values.

\begin{example}   \ (ex. \ref{ex:dpexample2} and \ref{ex:motivcex2} cont.) \ The extension for the $\nit{Access}$ view, given by query
 (\ref{eq:access}), is as follows:

\vspace{2mm}
\begin{multicols}{2}
{\small
\begin{tabular}{l|c|c|} \hline
$\nit{Access}(D)$ & \nit{User} & \nit{File} \\\hline
 & {\sf Joe}  & $f_1$ \\
  & {\sf Joe}  & $f_2$\\
& \sf{Tom} & $f_1$\\
& \sf{Tom} & $f_2$\\
& \sf{Tom} & $f_3$\\
& {\sf John} & $f_1$\\
& {\sf John} & $f_3$\\
 \hhline{~--}
 \end{tabular}
}

\vspace{2mm}
\noindent Assume the access of {\sf Joe} to file $f_1$ \ -corresponding to the query answer $\langle {\sf Joe}, f_1 \rangle$- \ is deemed to be unauthorized, while all other users' accesses are considered to be authorized, i.e. the other  answers to the query are considered to be correct.
\end{multicols}

First, $\nit{GroupUser}({\sf Joe}, g_1)$ is a counterfactual cause for  answer $\langle {\sf Joe}, f_1 \rangle$, and then also an
actual cause, with empty contingency set. Now we are interested in causes for the answer $\langle {\sf Joe}, f_1 \rangle$ that keep all the other answers untouched.
$\nit{GroupUser}({\sf Joe}, g_1)$ is also a vcc-cause.

In fact, \
$\nit{Access}(D \smallsetminus \{\nit{GroupUser}({\sf Joe}, g_1)\}) = \nit{Access}(D) \smallsetminus \{\langle {\sf Joe}, f_1 \rangle\}$,
showing that after the removal of  $\nit{GroupUser}({\sf Joe}, g_1)$, \red{all
the other previous answers, i.e. those in the table $\nit{Access}(D)$ above and different from $\langle {\sf Joe}, f_1 \rangle$, remain}. So, $\nit{GroupUser}({\sf Joe}, g_1)$ is a vc-cause with empty contingency set, or equivalently, a vcc-cause.

$\nit{GroupFile(f_1, g_1})$ is also an actual cause for $\langle {\sf Joe}, f_1 \rangle$, actually a counterfactual cause. However, it is not a vcc-cause, because its removal
leads to the elimination of the previous answer $\langle {\sf John}, f_1 \rangle$. Even less could it be a vc-cause, because deleting a non-empty contingency set
together with  $\nit{GroupFile(f_1, g_1})$ can only make things worse: answer $\langle {\sf John}, f_1 \rangle$ would still be lost.

Actually, $\nit{GroupUser}({\sf Joe}, g_1)$ is the only vc-cause and the only vcc-cause for $\langle {\sf Joe}, f_1 \rangle$.

Let us assume that, instead of $D$, we have instance $D'$, with  extensions:

\vspace{0.3cm}
\begin{center}
{\small
\begin{tabular}{l|c|c|} \hline
\nit{GroupUser}' & \ \nit{User} \ & \nit{Group} \\\hline
 & {\sf Joe}  & $g_0$\\
 & {\sf Joe}  & $g_1$\\
 & {\sf Joe}  & $g_2$\\
& {\sf John} & $g_1$\\
& \sf{Tom} & $g_2$\\
& \sf{Tom} & $g_3$\\
& {\sf John} & $g_3$\\
 \hhline{~--} \end{tabular} ~~~~~~~\begin{tabular}{l|c|c|} \hline
\nit{GroupFiles}'  & \nit{File} & \nit{Group} \\\hline
& $f_1$ & $g_0$ \\
 & $f_1$ & $g_1$ \\
  & $f_1$ & $g_3$ \\
& $f_2$ & $g_2$\\
& $f_3$& $g_3$\\
 \hhline{~--}
\end{tabular}
}
\end{center}

\vspace{2mm} The answers to the query are the same as with $D$, in particular,
we still have $\langle {\sf Joe}, f_1 \rangle$ as an answer to the query.

 \red{With the modified instance, } $\nit{GroupUser({\sf Joe}, g_1})$ is not a counterfactual cause for $\langle {\sf Joe}, f_1 \rangle$ anymore, since this answer can still be obtained via the
tuples involving $g_0$.  However, $\nit{GroupUser({\sf Joe}, g_1})$ is an actual cause, with minimal contingency sets: $\Gamma_1 = \{\nit{GroupUsers}({\sf Joe},g_0) \}$
and $\Gamma_2 = \{\nit{GroupFiles}(f_1,g_0) \}$.

$\nit{GroupUser({\sf Joe}, g_1})$ is not a vcc-cause \red{any longer}, but it is a vc-cause, with minimal contingence sets $\Gamma_1$ and $\Gamma_2$ as above: Removing
$\Gamma_1$ or $\Gamma_2$ from $D'$ keeps  $\langle {\sf Joe}, f_1 \rangle$ as an answer. However,
both under $D' \smallsetminus (\Gamma_1 \cup \{\nit{GroupUser({\sf Joe}, g_1})\})$ and $D' \smallsetminus (\Gamma_2 \cup \{\nit{GroupUser({\sf Joe}, g_1})\})$
the answer $\langle {\sf Joe}, f_1 \rangle$ is lost, but
the other answers stay.
\boxtheorem
\end{example}

\begin{example}\label{ex:vcex1} \re{(ex. \ref{ex:cfex1} and \ref{ex:motvcex1} cont.)} The answer $\langle \nit{\sf John},{\sf XML}\rangle$  does not have any vc-cause. In fact, consider for example the tuple \nit{Author({\sf John}, {\sf TODS})} that is an actual cause for $\langle \nit{\sf John},{\sf XML}\rangle$, with two contingency sets, $\Gamma_1$ and $\Gamma_2$. It is easy to verify that none of these contingency sets satisfies the condition in Definition \ref{def:vccause}(b). For example,  the original answer $\langle \nit{{\sf John}, {\sf CUBE}}\rangle$ is not preserved in $D \smallsetminus \Gamma_1$. The same argument can be applied to all actual causes for $\langle \nit{\sf John},{\sf XML}\rangle$. \boxtheorem
\end{example}

Notice that  Definition \ref{def:vccause} could be generalized by considering that several answers are unexpected and the others are correct. This generalization can only affect the admissible contingency sets.

The notions of {vc-}causality and vc-responsibility have corresponding decisions problems, which can be defined in terms similar to those
for plain causality and responsibility.

\begin{definition}  \em \label{def:mracpV}
(a) The {\em  {vc-}causality decision problem} ({VCDP}) is about membership of \
$\mc{VCDP}(\mc{Q})$ $=\{(D, \bar{a}, \tau)~|~  \bar{a} \in \mc{Q}(D) \ \mbox{and } \tau \in \nit{vc\mbox{-}\!Causes}(D, \mc{Q}(\bar{a}))   \ \}$.

\noindent (b) The {\em  {vc-}causal responsibility decision problem} is about membership of:

$\mathcal{VRDP}(\mc{Q}) \ = \ \{(D,  \bar{a},\tau, v)~|~ \tau \in D^n, v \in \{0\} \ \cup \{\frac{1}{k}~|~k \in \mathbb{N}^+\}, $  $D \models \mc{Q}(\bar{a}),$\\ \hspace*{8cm} and \ $\nit{vc\mbox{-}}\rho_{_{\!\mc{Q}\!}}^D(\tau) > v  \}$. \boxtheorem
\end{definition}

Leaving the answers to a view fixed when finding causes for a query answer is a strong condition. Actually, as Example \ref{ex:vcex1} shows, sometimes there are no
vc-causes. For this reason it makes sense to study the complexity of deciding whether a query answer has a  {vc-}cause or not.  This is a relevant problem.  For illustration,
consider the query $\nit{Access}$ in Example \ref{ex:motivcex2}. The existence of a  {vc-}cause for an unexpected answer (unauthenticated access) to this query, tells us that it is possible to revoke the unauthenticated access without restricting other users' access permissions. 

\begin{definition} \em   \label{def:VCEP}
  For a monotone query $\mc{Q}$, the  {\em  {vc-}cause existence  problem} ({VCEP}) is (deciding about membership of):

$\mc{VCEP}(\mc{Q})$ $=\{(D, \bar{a})~|~  \bar{a} \in \mc{Q}(D) \ \mbox{and } \nit{vc\mbox{-}\!Causes}(D, \re{\mc{Q}(\bar{a})}) \not = \emptyset \ \}.$
\boxtheorem
\end{definition}

\subsection{Characterization of vc-causality}\label{sec:vcVsCaus}

In this section we  establish mutual reductions between the  delete-propagation problem and  view-conditioned QA-causality. They will be used in Section
 \ref{sec:dpcompx} to obtain some complexity results for view-conditioned causality.

\red{Next,} we show that, in order to check if there exists a solution to the view-side-effect-free deletion-problem for  \re{$\bar{a} \in \mc{V}(D)$} (cf. Definition \ref{def:FreeVDP}), it is good enough to check if $\bar{a}$ has a view-conditioned cause for $\bar{a}$.\footnote{ \ Since this delete-propagation problem does not explicitly involve anything like contingency sets, the existential problem in Definition \ref{def:FreeVDP}(b) is the right one to consider.}

\begin{proposition} \label{pro:VC&view} \em
  For an instance $D$ and a view  defined by a monotone query $\mc{Q}$, with $\bar{a} \in \mc{Q}(D)$, \ $(D, \bar{a}) \in \mc{VSEFP}(\mc{Q})$ \ iff \
 $\re{\nit{vc\mbox{-}\!Causes}(D, \mc{Q}(\bar{a})) \not = \emptyset}$.
\end{proposition}

\dproof{Assume $\bar{a}_1$ has a view-conditioned cause $\tau$. According to
Definition \ref{def:vccause}, there exists a $\Gamma \subseteq D$, such that $D \smallsetminus (\Gamma \cup \{\tau\}) \not \models \mc{Q}(\bar{a})$, $D \smallsetminus \Gamma \models \mc{Q}(\bar{a})$, and  $D \smallsetminus (\Gamma \cup \{\tau\}) \models \mc{Q}(\bar{a}')$, for every \ $\bar{a}' \in \mc{Q}(D)$ with  $\bar{a}' \neq \bar{a}$. So, $\Gamma \cup \{\tau\}$ is a view-side-effect-free delete-propagation solution for $\bar{a}$; and $(D, \bar{a}) \in \mc{VSEFP}(\mc{Q})$. A similar argument applies in the other direction.
}

\begin{example}\label{ex:Vpex2} (ex. \ref{ex:dpp}, \ref{ex:Vpex1} and \ref{ex:vcex1} cont.) We obtained in Example \ref{ex:dpp}(b) that there is no view-side-effect-free solution to the delete-propagation problem for
the view tuple $\langle\nit{\sf John},{\sf XML}\rangle$. This coincides with the result  in  Example  \ref{ex:vcex1}, and confirms Proposition \ref{pro:VC&view}.
\boxtheorem
\end{example}

Next, we show that  {vc-}causes for an answer can \bblue{be} obtained from solutions to a corresponding view-side-effect-free deletion-problem.
\begin{proposition} \label{pro:VCcausefromview} \em
For an instance $D=D^n \cup D^x$ and a monotone query $\mc{Q}(\bar{x})$ with $\bar{a} \in \mc{Q}(D)$,  $\tau \in D^n$ is a
 {vc-}cause for $\bar{a}$  \ iff \ there is  $D' \subseteq D$, with $\tau \in
(D\smallsetminus D') \subseteq D^n$, that  is a solution to the view-side-effect-free deletion-problem for $\bar{a}$.
\end{proposition}

\dproof{ Similar to the proof of Proposition \ref{pro:causefromview}.}

\subsection{Complexity of vc-causality} \label{sec:dpcompx}

\red{In the following, we}  investigate the complexity of the view-conditioned causality problem  (cf. Definition \ref{def:mracpV}). For this, we take advantage of the connection between  {vc-}causality and view-side-effect-free  delete-propagation.

First, the following result about  the {\em {vc-}cause existence problem} (cf. Definition \ref{def:VCEP}) is obtained from the {\em NP}-completeness of the view-side-effect-free delete-propagation decision problem
for conjunctive views \cite[theorem 2.1]{BunemanKT02} and Proposition \ref{pro:VC&view}.

\begin{proposition} \label{pro:VCcauseexistence} \em
For  {CQ}s $\mc{Q}$, $\mc{VCEP}(\mc{Q})$ is \nit{ {NP}}-complete (in data).
\end{proposition}

\dproof{For membership of {\em NP}, the following is a non-deterministic \nit{PTIME} algorithm for $\mc{VCEP}$: Given $D$ and answer $\bar{a}$ to $\mc{Q}$, guess a subset $\Gamma \subseteq D^n$ and a tuple $\tau \in D^n$, return {\em yes} if $\tau$ is a vc-cause for $\bar{a}$ with contingency set $\Gamma$; otherwise return {\em no}. This test can be performed in \nit{PTIME} in the size of $D$.

Hardness is by the reduction from the (\nit{NP}-hard) view-side-effect-free delete-propagation problem  that is  explicitly given  in the formulation of Proposition \ref{pro:VC&view}.}

The next result is about deciding {vc-}causality (cf. Definition \ref{def:mracpV}).

\begin{proposition} \label{pro:VCcausecausality} \em
For  {CQ}s $\mc{Q}$, $\mc{VCDP}(\mc{Q})$ is  \nit{ {NP}}-complete (in data).
\end{proposition}

\dproof{{\em Membership}: \  For an input $(D, \bar{a})$, non-deterministically guess $\tau \in D^n$ and $\Gamma \subseteq D^n$, with $\tau \notin \Gamma$. If $\tau$ is a vc-cause for $\bar{a}$ with contingency set $\Gamma$ (which can be checked in polynomial time), return {\em yes}; otherwise return {\em no}.

{\em Hardness:} \ Given an instance $D$ and $\bar{a} \in \mc{Q}(D)$, it is easy to see that: \ $(D, \bar{a}) \in {\mc{VCEP}}(\mc{Q})$ \ iff \ there is ${\tau} \in D^n$ with $(D, \bar{a}, {\tau}) \in {\mc{VCDP}}(\mc{Q})$.
This immediately gives us a one-to-many reduction from $\mc{VCEP}(\mc{Q})$: \ $(D, \bar{a})$ is mapped to the polynomially-many inputs of the form $(D, \bar{a}, {\tau})$ for ${\mc{VCDP}}(\mc{Q})$, with $\tau \in D^n$.
The answer for  $(D, \bar{a})$ is {\em yes} iff at least for one $\tau$, $(D, \bar{a}, {\tau})$ gets answer {\em yes}. This is a polynomial number of membership tests for ${\mc{VCDP}}(\mc{Q})$.
}

\vspace{2mm}
In this result, $\nit{NP}$-hardness is defined in terms of ``Cook (or Turing) reductions" as opposed to many-one (or Karp) reductions \cite{garey,goldreich}.
$\nit{NP}$-hardness
under many-one reductions implies $\nit{NP}$-hardness under Cook
reductions, but the converse, although  conjectured
not to hold, is an open problem. However, for Cook
reductions, it is still true that there is no efficient algorithm
for an $\nit{NP}$-hard problem, unless $P = \nit{NP}$.

Finally, we settle the complexity of the  {vc-}causality responsibility problem for conjunctive queries.

\begin{proposition} \label{pro:VCcauseresponsibility} \em
For  {CQ}s $\mc{Q}$, $\mc{VRDP}(\mc{Q})$ is  \nit{ {NP}}-complete \ (in data).
\end{proposition}

\dproof{{\em Membership:} \ For an input $(D,\bar{a},\tau,v)$, non-deterministically guess $\Gamma \subseteq D^n$, and return {\em yes} if  $\tau$ is a vc-cause for $\bar{a}$ with contingency set
$\Gamma$, and
$|\Gamma| < \frac{1}{v}$. Otherwise, return {\em no}. The verification can be done in \nit{PTIME} in data.

{\em Hardness:} \ By reduction from the VCDP problem, shown to be $\nit{NP}$-complete
in Proposition \ref{pro:VCcausecausality}.

Map $(D,\bar{a},\tau)$, an input for  $\mc{VCDP}(\mc{Q})$, to  the input $(D, \bar{a},\tau, k)$ for $\mc{VRDP}(\mc{Q})$, where $k=\frac{1}{|D|+1}$. Clearly,
$(D, \bar{a}, \tau) \in \mc{VCDP}(\mc{Q})$ iff $(D, \tau, \bar{a}, k) \in \mathcal{VRDP}(\mc{Q})$.
This follows from the fact  that $\tau \in D^n$ is an actual cause for $\bar{a}$ iff $\nit{vc}\mbox{-}\rho_{_{\!\mc{Q}(\bar{a})\!}}^D(\tau) \geq \frac{1}{|D|}$.  
}

Notice that the previous proof uses a Karp reduction, but from a problem identified as $\nit{NP}$-hard through the use of a Cook reduction (in Proposition \ref{pro:VCcausecausality}).

All results on vc-causality in this section also hold for  {UCQ}s.

\ignore{
As mentioned in Section \ref{sec:causeintro}, causal responsibility computation (more precisely the  {RDP} problem in Definition \ref{def:resp}) is tractable for weakly linear queries. We can take advantage of this result and obtain,
via Proposition \ref{pro:causeminSDP}, a new tractability result for the minimum-source-side-effect deletion-problem, which has been shown to be \nit{ {NP}}-hard  for general  {CQ}s in \cite{BunemanKT02}.

\begin{proposition} \label{pro:dichsse} \em
For weakly linear queries, the minimum-source-side-effect decision problem is tractable.\boxtheorem
\end{proposition}

The class of weakly linear queries generalizes that of linear queries (cf. Section \ref{sec:causeintro}). So, Proposition \ref{pro:dichsse} also holds for linear queries.

In \cite{BunemanKT02} it has been shown that the minimum-source-side-effect decision problem is tractable for the class of
project-join queries with {\em  chain joins}. Now, a join on $k$ atoms with different predicates, say
$R_1,..., R_k$, is a chain join if there are no  attributes (variables) shared by
any two atoms $R_i$ and $R_j$ with $j > i + 1$. That is, only consecutive relations may share attributes. For example, $\exists xvyu(A(x) \wedge S_1(x, v) \wedge S_2(v, y) \wedge R(y, u) \wedge S_3(y, z))$ is a project-join query with chain joins.

We observe that project-join queries with chain joins correspond
linear queries. Actually, the tractability results for these classes of queries are both obtained via a reduction to maximum flow problem \cite{Meliou2010a, BunemanKT02}.  As a consequence, the result in Proposition \ref{pro:dichsse} extends that in \cite{BunemanKT02}, from linear queries to weakly-linear queries. For example, $\exists xyz(R(x, y) \wedge S(y, z) \wedge T(z, x) \wedge V(x))$ is not linear (then, nor with chain joins), but it is weakly linear \cite{Meliou2010a}.
}

\section{QA-Causality under Integrity Constraints} \label{sec:c&ic}

We start with some observations and examples on  QA-causality in the presence of integrity constraints (ICs). First, at the basis of Halpern \& Pearl's  approach to causality \cite{Halpern05},
we find {\em interventions}, i.e.
actions on the model that determine counterfactual scenarios. In  databases, they take the
form of database updates,  in particular, tuple deletions, which is the scenario we have consider so far. Accordingly, if a database $D$ is expected to satisfy a given set of integrity constraints (that should also be considered as parts of the ``model"), the
instances obtained from $D$ by tuple deletions (as interventions), as used to determine causes, should  also satisfy the
ICs.

On a different side, QA-causality  as introduced in \cite{Meliou2010a}  is {\em insensitive} to equivalent query
rewriting (as first pointed out in \cite{Glavic11}): On the same instance, causes for query answers coincide
for logically equivalent queries. \ignore{ This property is important since databases engines are accustomed to choose and evaluate the simplest equivalent query to the
one at hand. The fact that QA-causality posses this property guarantees that QA-causes given by this definition are the intended one.}
However, QA-causality might be sensitive to equivalent query rewritings in the presence of ICs, as the following example shows.

\begin{example}\label{ex:ICex1}
Consider a relational schema $\mc{S}$ with predicates $\nit{Dep(DName},$ $\nit{TStaff)}$ and
$\nit{Course(CName},$ $\nit{LName, DName)}$. Consider the instance $D$ for $\mc{S}$:

\begin{center}
\footnotesize
\begin{tabular}{c|c|c|} \hline
\nit{ Dep} & \nit{DName} &\nit{TStaff}  \\\hline
$t_1$& {\sf Computing} & {\sf John}   \\
$t_2$& {\sf Philosophy} &  {\sf Patrick}   \\
$t_3$&{\sf Math}  &  {\sf Kevin}   \\
 \hhline{~--} \end{tabular}~~~~ 
 \begin{tabular}{c|c|c|c|} \hline
\nit{Course}  & \nit{CName} & \nit{TStaff} & \nit{DName} \\\hline
$t_4$&{\sf COM08} & {\sf John}  & {\sf Computing} \\
$t_5$&{\sf Math01} & {\sf Kevin}  & {\sf Math} \\
$t_6$&{\sf HIST02}&  {\sf Patrick}   &{\sf Philosophy} \\
$t_7$&{\sf Math08}&  {\sf Eli}   &{\sf Math}  \\
$t_8$&{\sf COM01}&  {\sf John} &{\sf Computing} \\
 \hhline{~---}
\end{tabular}
 \end{center}

\noindent where all the tuples are endogenous. Now, consider the  CQ, $\mc{Q}$, that collects the teaching staff who are
lecturing  in the department they are associated with:
 \begin{eqnarray}
\!\!\!\nit{Ans}_{\mc{Q}}(\nit{TStaff})&\leftarrow&\nit{Dep(DName,TStaff}),   \label{eq:heads}\\
&&~~~~~ \nit{Course(CName,TStaff, DName}). \nonumber
\end{eqnarray}
The answers are: \ $\mc{Q}(D) = \{{\sf John}, {\sf Patrick}, {\sf Kevin}\}$. Answer $\langle {\sf John} \rangle$ has the following actual causes: $t_1$, $t_4$ and $t_8$.
$t_1$ is a counterfactual cause, $t_4$ has a single minimal contingency set $\Gamma_1=\{t_8\}$; and
 $t_8$ has a single minimal contingency set $\Gamma_2=\{t_4\}$.

Now, consider the following inclusion dependency that is satisfied by $D$:
\begin{equation}
\psi\!: \ \ \forall x \forall y \ (\nit{Dep}(x, y) \rightarrow \exists u  \  \nit{Course}(u, y, x)). \label{eq:ind}
\end{equation}
In the presence of $\psi$, \ $\mc{Q}$ is equivalent to the  query $\mc{Q}'$ given by:
  \begin{eqnarray}
\nit{Ans}_{\mc{Q}'}(\nit{TStaff}) &\leftarrow& \nit{Dep(DName,TStaff})).  \label{eq:heads2}
\end{eqnarray}
That is, \ $\mc{Q} \equiv_{\{\psi\}}\mc{Q}'$.

For query $\mc{Q'}$, $\langle {\sf John} \rangle$ is still an answer from $D$. However, considering only query $\mc{Q}'$ and instance $D$, this answer has
a single cause, $t_1$, which is also a counterfactual cause. The question is whether $t_4$ and $t_8$ should still be considered as causes for answer $\langle {\sf John} \rangle$
in the presence of $\psi$.

\re{Now consider the query $\mc{Q}_1$ given by
\begin{eqnarray}
\nit{Ans}_{\mc{Q}_1}(\nit{TStaff}) &\leftarrow& \nit{Course(CName,TStaff, DName}).  \label{eq:heads3}
\end{eqnarray}
$\langle {\sf John} \rangle$ is an answer, and $t_4$ and $t_8$ are the only actual causes, with contingency sets $\Gamma_1 = \{t_8\}$ and $\Gamma_2 = \{t_4\}$, resp. }

\re{In the presence of
$\psi$, one should wonder if also $t_1$ would be a cause (it contains the referring value {\sf John} in table $\nit{Dept}$), or, if not, whether its presence would make the previous causes less responsible. }
\boxtheorem
\end{example}
\begin{definition} \label{def:causeIC} \em
Given an instance $D=D^n \cup D^x$ that satisfies a set $\Sigma$ of ICs, i.e. $D\models \Sigma$, and a
monotone query $\mc{Q}$ with $D \models \mc{Q}(\bar{a})$, a tuple $\tau \in D^n$ is an  {\em actual cause for  $\bar{a}$
under $\Sigma$} \ if there is $\Gamma \subseteq D^n$, such that:
\begin{itemize}
\item[(a)]
$ D \smallsetminus \Gamma  \models  \mc{Q}(\bar{a})$, \ and \ (b) \
$ D \smallsetminus \Gamma \models \Sigma$.
\item[(c)] $ D \smallsetminus (\Gamma \cup \{t\}) \not \models  \mc{Q}(\bar{a})$, \ and \ (d) \ $ D \smallsetminus (\Gamma \cup \{t\}) \models \Sigma$.

\end{itemize}
$\nit{Causes}(D, \mc{Q}(\bar{a}), \Sigma)$
denotes the set of actual causes for $\bar{a}$ under $\Sigma$.  \re{For $\tau \in \nit{Causes}(D, \mc{Q}(\bar{a}), \Sigma)$,  $\nit{Cont}(D, \mc{Q}(\bar{a}), \tau,\Sigma)$ and $\nit{Cont}^s(D, \mc{Q}(\bar{a}), \tau,\Sigma)$ denote
the set of contingency sets, resp. subset-minimal contingency sets, for $\tau$ under $\Sigma$.} \boxtheorem
\end{definition}

\re{The {\em responsibility} of $\tau$ as a cause for an answer $\bar{a}$ to query $\mc{Q}$ under a set $\Sigma$
 of ICs, denoted by $\rho_{_{\!\mc{Q}(\bar{a})\!}}^{D,\Sigma}(\tau)$, is defined exactly as in Section  \ref{sec:causeintro}.}

\begin{example}\label{ex:ICex2} (ex. \ref{ex:ICex1} cont.)  Consider  query $\mc{Q}$ in (\ref{eq:heads}), and its answer  $\langle {\sf John} \rangle$.
Without the constraint $\psi$ in (\ref{eq:ind}), tuple $t_4$ was a cause with minimal contingency set $\Gamma_1=\{t_8\}$.

Now, it holds $D \smallsetminus \Gamma_1 \models \psi$, but
$D \smallsetminus \red{(\Gamma_1 \cup \{t_4\})} \not \models \psi$. So, in presence of $\psi$, and applying Definition \ref{def:causeIC},
$t_4$ is not longer an actual cause for answer $\langle {\sf John} \rangle$. \re{The same happens with $t_8$. However, $t_1$ is still an actual (counterfactual) cause, and the only one. So, it holds: \
$\nit{Cause}(D, \mc{Q}({\sf John}), \psi) \subsetneqq
 \nit{Causes}(D, \mc{Q}({\sf John}))$.}

\re{Notice that
$\mc{Q}$ and $\mc{Q'}$ in (\ref{eq:heads2}) have the same actual causes for answer $\langle {\sf John} \rangle$ under $\psi$, namely $t_1$.}

\re{Now consider query $\mc{Q}_1$ in (\ref{eq:heads3}), and its answer $\langle {\sf John} \rangle$. Tuples $t_4$ and $t_8$ are still (non-counterfactual) actual causes in the presence of
$\psi$. However, their previous contingency sets are not such anymore: \ $D  \smallsetminus (\Gamma_1 \cup \{ t_4\}) \not \models \psi$, \ $D  \smallsetminus (\Gamma_2 \cup \{ t_8\}) \not \models \psi$.
Actually, the smallest contingency set for $t_4$ is $\Gamma_3 = \{t_8, t_1\}$; and for $t_8$, $\Gamma_4 = \{t_4, t_1\}$. Accordingly, the causal responsibilities of $t_4, t_8$ decrease in the presence of
$\psi$: \  $\rho_{_{\mc{Q}({\sf John})}}^D(t_4) = \frac{1}{2}$, but  $\rho_{_{\mc{Q}({\sf John})}}^{D,\psi}(t_4) =\frac{1}{3}$.}

\re{In the presence of $\psi$, tuple $t_1$ is still not an actual cause for answer $\langle {\sf John} \rangle$ to $\mc{Q}_1$. For example, if we check the conditions in Definition \ref{def:causeIC}, with  $\Gamma_1$ as potential
contingency set, we find that  \ (a),(b) and (d) hold: \
$ D \smallsetminus \Gamma_1  \models  \mc{Q}_1({\sf John})$, \
$ D \smallsetminus \Gamma_1 \models \varphi$, \ and $ D \smallsetminus (\Gamma_1 \cup \{t_1\}) \models \psi$, resp. However, (c) does {\em not} hold: \ $D \smallsetminus (\Gamma_1 \cup \{t_1\})  \models  \mc{Q}_1({\sf John})$.
For any other potential contingency set, some of the conditions (a)-(d) are not satisfied.}
\boxtheorem
\end{example}

Functional dependencies (FDs) are never violated by tuple deletions. For these reason, conditions (b) and (d) in Definition \ref{def:causeIC}, those that have to do with the ICs, are always satisfied. So, FDs \red{should
have no effect on} the set of causes
for a query answer. Actually, this applies to the more general class of {\em denial constraints} (DCs), i.e. of the form
$\neg \forall \bar{x}(A_1(\bar{x}_1) \wedge \cdots \wedge A_n(\bar{x}_n))$, with $A_i$ a database predicate or a built-in.

\red{More general ICs may make sets of causes grow, and also the sizes of minimal contingency sets. Accordingly the responsibilities of causes may decrease. This is in line with tuple dependencies captured by ICs. For example, the satisfaction of tgds may force additional tuple deletions, those appearing in their antecedents (cf. Example \ref{ex:ICex2}). Intuitively, the responsibility is spread out through tuple dependencies.}

\vspace{-1mm}\begin{proposition} \em
Consider an instance $D$, a monotone query $\mc{Q}$, and a set of ICs $\Sigma$, \re{such that $D \models \Sigma$}. The following hold:
\begin{enumerate}[(a)]
\item  $\nit{Causes}(D, \mc{Q}(\bar{a}), \Sigma) \subseteq
 \nit{Causes}(D, \mc{Q}(\bar{a}))$. \  \re{Furthermore, for every $\tau \in D$, \ $\rho_{_{\mc{Q}(\bar{a})}}^{D,\Sigma}(\tau) \ \leq \ \rho_{_{\mc{Q}(\bar{a})}}^D(\tau)$.}

 \item $\nit{Causes}(D, \mc{Q}(\bar{a}), \emptyset) = \nit{Causes}(D, \mc{Q}(\bar{a}))$.

 \item If $\Sigma$ is a set  of DCs,
 $\nit{Causes}(D, \mc{Q}(\bar{a}), \Sigma) = \nit{Causes}(D, \mc{Q}(\bar{a}))$. \re{Furthermore, for every $\tau \in D$, \ $\rho_{_{\mc{Q}(\bar{a})}}^{D,\Sigma}(\tau) = \rho_{_{\mc{Q}(\bar{a})}}^D(\tau)$.}

  \item For a monotone  query $\mc{Q}'$ with
  $\mc{Q}' \equiv_\Sigma \mc{Q}$, it holds \
  $\nit{Causes}(D, \mc{Q}(\bar{a}), \Sigma) = \nit{Causes}(D, \mc{Q}'(\bar{a}),\Sigma)$.\boxtheorem
\end{enumerate}
\ignore{  \item For a monotone query $\mc{Q}'$ with  $\mc{Q}' \subseteq \mc{Q}$ and
  $\mc{Q}' \equiv_\Sigma \mc{Q}$,\ignore{\footnote{ \ This means $\mc{Q}' \subseteq \mc{Q}$ and there is no  $\mc{Q}''$ with
  $\mc{Q}'' \subsetneqq \mc{Q}'$  and $\mc{Q}'' \equiv_\Sigma \mc{Q}$. }} \ it holds \
  $\nit{Causes}(D, \mc{Q}(\bar{a}), \Sigma) = \nit{Causes}(D, \mc{Q}'(\bar{a})).$ \boxtheorem}
\end{proposition}

\dproof{(a) \ Any contingency set $\Gamma$ used for $\tau \in \nit{Causes}(D, \mc{Q}(\bar{a}), \Sigma)$,   can be used as a contingency
set for the definition of causality without ICs (which are those in Definition \ref{def:causeIC}(a,c)): \ $\nit{Cont}(D, \mc{Q}(\bar{a}), \tau,\Sigma)
\subseteq \nit{Cont}(D, \mc{Q}(\bar{a}), \tau)$. The same inclusion holds for subset-minimal contingency sets.

\noindent (b) \ For every contingency set $\Gamma$ for a cause $\tau$ without ICs, conditions in Definition \ref{def:causeIC}(b,d)
are trivially satisfied with an empty set of ICs.

\noindent (c) \ When $D \models \Sigma$, and $\Sigma$ are DCs, every subset of $D$ also satisfies $\Sigma$. Then, the new conditions on candidate contingency sets, those in Definition \ref{def:causeIC}(b,d),
are immediately satisfied. Since the same contingency sets apply both with or without ICs, the responsibility does not change.

\noindent (d) \ For  a potential cause $\tau$ with a candidate contingency set $\Gamma$, conditions in  Definition \ref{def:causeIC}(a,c) will be always simultaneously satisfied
for $\mc{Q}$ and $\mc{Q}'$, because according to the conditions in  Definition \ref{def:causeIC}(b,d), both $D \smallsetminus \Gamma$ and $D \smallsetminus (\Gamma \cup \{\tau\})$ satisfy $\Sigma$.
}

Notice that Example \ref{ex:ICex2} shows that the inclusion in item (a) above can be proper. It also shows that for a same actual cause, with and without ICs, the inequality of responsibilities may be strict.

Item (d) above corresponds to the equivalent rewriting of the query in (\ref{eq:heads}) into query (\ref{eq:heads2}) under the referential constraints. As shown in Example \ref{ex:ICex2},
under the latter both queries have the same causes.\footnote{ Notice that this rewriting  resembles the resolution-based rewritings used
in {\em semantic query optimization} \cite{minker}.}
\re{The monotonicity condition on $\mc{Q}'$ in item (d) is necessary, first to apply the notion of cause to it, but more importantly, because monotonicity is not implied by the monotonicity of $\mc{Q}$ and
query equivalence under $\Sigma$. In fact, for schema $\mc{S} = \{R(A,B), S(A,B)\}$, the FD \ $R\!:A \rightarrow B$, the BCQ $\mc{Q}\!: \ \exists x \exists y \exists z(R(x,y) \wedge R(x,z) \wedge y \neq z)$, and
the non-monotonic Boolean  query $\mc{Q}'\!: \ \exists x \exists y \exists z(R(x,y) \wedge R(x,z) \wedge \neg S(x,y) \wedge y \neq z)$, it holds $\mc{Q} \equiv_{\mbox{FD}} \mc{Q'}$.}

{\em All the causality-related decision and computational problems for the case without ICs can be easily redefined in the presence of a set $\Sigma$ of ICs, that we now make explicitly appear as a problem parameter, such as in
$\mc{RDP}(\mc{Q},\Sigma)$, for the responsibility decision problem.}

Since {FD}s have no effect on causes, the causality-related decision problems in the presence of  {FD}s have the
same complexity upper bound as causality without FDs. For example, for  a set $\Sigma$ of FDs, $\mc{RDP}(\mc{Q},\Sigma)$, the responsibility problem now under
FDs, is \nit{NP}-complete, since this is already the case without ICs \cite{Meliou2010a}.

When an instance satisfies
a set of  {FD}s, the decision problems may become tractable depending on the query structure. A particular syntactic class of CQs is that of
  key-preserving  CQs: Given a set $\kappa$ of key constraints (KCs), a CQ $\mc{Q}$ is {\em key-preserving} (more precisely, {\em $\kappa$-preserving}) if
\re{the key attributes
of the
relations appearing in $\mc{Q}$ are all included among the non-existentially quantified
attributes of $\mc{Q}$ \ \cite{Cong12}.} \re{For, example, for the schema $S(\underline{A},\underline{B},C), R(\underline{C},D)$, with the keys underlined, the queries
$\mc{Q}_1(y,z)\!:  \exists x S(x,y,z)$, \ $\mc{Q}_2(x,y,z)\!:   (S(x,y,z) \wedge R(z,v))$ are not key-preserving, but
$\mc{Q}_3(x,y)\!:  \exists z S(x,y,z)$ and $\mc{Q}_2^\prime(x,y,z,w,z)\!: \   (S(x,y,z) \wedge R(w,v) \wedge z = w)$ are.}
It turns out that, in the case of key-preserving CQs, deciding responsibility over instances that satisfy the key constraints (KCs) is in
{\nit{PTIME}} \cite{Cibele15}.

The view-side-effect-free delete propagation (VSEFD) problem can be easily reformulated in the presence of ICs, by including their satisfaction in Definition
\ref{def:FreeVDP}, both by $D$ and the instance resulting from delete propagation, $D \smallsetminus \Lambda$. Furthermore, the mutual characterizations between the VSEFD and
 view-conditioned causality problems of Section \ref{sec:vcVsCaus} still hold in the presence of ICs.

 It turns out that the decision version of  the view-side-effect-free deletion problem for key preserving CQs is tractable in data complexity \cite{Cong12}.  By appealing to the connection in Section \ref{sec:vcc} between vc-causality and that form
of delete-propagation,  vc-responsibility under KCs becomes tractable.\footnote{ \ Actually, the result in \cite{Cong12} just mentioned  holds for single tuple deletions (with multiple deletions it can be {\em NP}-hard), which is the case
in the causality setting, where a single answer is hypothetically deleted.} However, it is intractable in general, because the
problem without KCs already is, as shown in Proposition \ref{pro:VCcauseresponsibility}).


\begin{proposition} \label{pro:ICcause} \em
Given a set $\kappa$ of KCs, and a key-preserving {{CQ}} query  $\mc{Q}$, deciding $\mc{VRDP}(\mc{Q},\kappa)$ is in  {\nit{PTIME}}. \boxtheorem
\end{proposition}

\re{Other classes of (view-defining) CQs for which different variants of delete-propagation are tractable are investigated
in \cite{Kimelfeld12a,Kimelfeld12b} (generalizing  those in \cite{Cong12}). The connections between
delete-propagation and causality established in Sections \ref{sec:delp&cause} and \ref{sec:vcc} should allow us to obtain new tractability results for causality.}

Our next result tells us that it is possible to capture vc-causality
through non-conditioned QA-causality under tuple-generating dependencies (tgds).

\vspace{-1mm}
\begin{lemma} \label{pro:vc-ICs} \em   For every  instance $D$ an for a schema $\mc{S}$,
$\mc{Q}(\bar{x}) \in \mc{L}(\mc{S})$  a conjunctive query with $n$ free variables, and $\bar{a} \in \mc{Q}(D)$, there
 is a \re{tgd} $\psi$  over schema $\mc{S} \cup \{V\}$, with $V$ a fresh $n$-ary predicate, and an instance $D'$ for $\mc{S} \cup \{V\}$, such that
$\nit{vc\mbox{-}\!Causes}(D, \mc{Q}(\bar{a}))=\nit{Causes}(D', \mc{Q}(\bar{a}), \{\psi\})$. \boxtheorem
\end{lemma}

\hproof{Consider the instance $D' := D \cup \{V(\bar{c})~|~ \bar{c} \in (\mc{Q}(D)\smallsetminus\{\bar{a}\})$, where the second disjunct is the extension for predicate $V$. The  tgd
 $\psi$ over schema $\mc{S} \cup \{V\}$ is \ $\forall \bar{x}(V(\bar{x}) \rightarrow \mc{Q}(\bar{x}))$. }

In the absence of ICs, deciding causality for CQs  is tractable \cite{Meliou2010a}, but their presence may have an
impact on this problem.

\begin{proposition} \label{pro:keyp} \em
For  a CQ $\mc{Q}$ and \re{a tgd $\psi$},
  $\mc{CDP}(\mc{Q},\{\psi\})$ is {\em  {NP}}-complete. \boxtheorem
\end{proposition}

\hproof{Membership is clear. Hardness is established by reduction from the \nit{NP}-complete vc-causality decision problem (cf. Proposition \ref{pro:VCcausecausality}) for a  CQ $\mc{Q}(\bar{x})$ over schema $\mc{S}$.
Now, consider the schema $\mc{S}':= \mc{S} \cup \{V\}$ and the  tgd $\psi$ as in Lemma \ref{pro:vc-ICs}. In order to decide about $(D,\mc{Q}(\bar{a}), \tau)$'s membership of $\mc{VCDP}(\mc{Q})$,
consider the instance $D'$ for $\mc{S}'$ as in Proposition \ref{pro:vc-ICs}. It holds: \ $(D,\mc{Q}(\bar{a}), \tau) \in \mc{VCDP}(\mc{Q})$ iff
$(D',\mc{Q}(\bar{a}), \tau) \in \mc{CDP}(\mc{Q},\{\psi\})$.}

\subsection{\re{Causality under ICs via view-updates and abduction}}\label{sec:abdIC}

In this work we have connected QA-causality with both abduction and view-updates in form of delete-propagations. It is expected to find connections between
causality under ICs and those two other problems in the presence of ICs, as the following example suggests.

\begin{example} (ex. \ref{ex:ICex1} cont.) Formulated as an abduction problem, we have
the query $\mc{Q}$ specified in by the Datalog rule in (\ref{eq:heads}), defining an intentional predicate, $\nit{Ans}_{\mc{Q}}(\nit{TStaff})$.
All the tuples in the underlying database $D$, all  endogenous,
are  considered to be abducible. The view-update request is the deletion of $\langle {\sf John}\rangle$ from $\mc{Q}(D)$ (more precisely, from $\nit{Ans}_{\mc{Q}}(D)$). As an
abduction task, it is about giving an explanation for obtaining tuple $\nit{Ans}_{\mc{Q}}({\sf John})$.

According to  our approach to abduction of Section \ref{sec:abdandcause}, the abductive explanations are obtained from (and also lead to) maximal subsets $E$ of  $D$, such that
$E$ plus the query rule (\ref{eq:heads}) does not entail $\nit{Ans}_{\mc{Q}}({\sf John})$ anymore. These sets are: \ $E_1 = D \smallsetminus \{t_1\}$, and $E_2 = D \smallsetminus \{t_4,t_8\}$,
and are determined by finding minimal abductive explanations for $\nit{Ans}_{\mc{Q}}({\sf John})$. So far, all this without considering the IC $\psi$ in (\ref{eq:ind}).

Now, these maximal sub-instances  have to be examined at the light of the IC. In this case, $E_1$ does satisfy $\psi$, but $E_2$ does not. So, the
latter is rejected. As a consequence, the only admissible update is the deletion of $t_1$ from $D$, which coincides with having $t_1$ as the only actual cause
under the IC, as determined in Example \ref{ex:ICex2}.
 \boxtheorem
\end{example}
This example shows that, and how,  (minimal) abductive explanations, and also admissible view-updates, could be used to define, provide alternative characterizations,   and compute actual causes in the presence of ICs. In this case, and according to
Section \ref{sec:delp&cause}, an admissible view-update (under the ICs) should be in correspondence, by definition, with an {\em admissible}
combination
of an actual cause and one of its
contingency sets. This would make, in the previous example,  $t_1$ the only actual cause (also counterfactual) for $\langle {\sf John}\rangle$ under $\psi$, as expected from the direct definition of cause
under ICs.

Both view-updates and abduction can be defined in the presence of ICs. In particular, theories written in languages of logic programming have been considered
as underlying theories for abduction and view updates in the presence of ICs \cite{Kakas90,kowalski}. More specifically, in \cite{Console95}, view updates via abductive explanations are investigated  in the context of stratified logic programs
with ICs on the extensional database (as opposed to on the intentional relations).

We briefly
illustrate using our ongoing example how Datalog abduction \`a la logic programming with constraints \cite{kowalskiVLDB} could be used to determine causes in the presence of ICs.

\begin{example} (ex. \ref{ex:ICex1} and \ref{ex:ICex2} cont.) \label{ex:lp} Consider query $\mc{Q}_1$, defined by the Datalog rule in (\ref{eq:heads3}), and the IND $\psi$ in (\ref{eq:ind}). We want to compute
the causes for answer {\sf John} by applying a resolution-based refutation procedure that generates candidate causes, but checks possible support from ICs, for underlying causes:
\begin{eqnarray*}
&\leftarrow& \nit{Ans}_{\mc{Q}_1}({\sf John}) \hspace{1.3cm}\mbox{(negated answer)}\\
\nit{Ans}_{\mc{Q}_1}(x) &\leftarrow&\nit{Course}(u,x, y)\\
&\leftarrow&\nit{Course}(u,{\sf John}, y)\\
\nit{Course}({\sf COM08},{\sf John},{\sf Computing})&\leftarrow& \hspace{3.8cm}\mbox{(from $D$)~~(*)}\\
&\leftarrow&\hspace{2.9cm}\mbox{(tuple is candidate)}\\
\nit{Course}(u, {\sf John}, y) &\leftarrow& \nit{Dep}(y, {\sf John}) \hspace{0.7cm}\mbox{(check IND with (*))}\\
&\leftarrow& \nit{Dep}(y, {\sf John})\\
\nit{Dep}({\sf Computing}, {\sf John}) &\leftarrow& \hspace{4.5cm}\mbox{(from $D$)}\\
&\leftarrow& \hspace{0.9cm}\mbox{(tuple is candidate, no more IC)}
\end{eqnarray*}
The successful refutation shows $\nit{Dep}({\sf Computing}, {\sf John})$ as an abductive explanation (or a cause).\footnote{ \ More precisely,  a Skolem functional term $f(y, {\sf John})$ should replace variable $u$ in $\nit{Course}(u, {\sf John}, y) \leftarrow \nit{Dep}(y, {\sf John})$ \cite{lloyd87}.}

Notice that our additional checking above of (*) with the IND can be seen as generating a new query through the interaction
of (\ref{eq:heads3}) and the IND, namely: \  $\nit{Ans}_{\mc{Q}_1}'(x) \leftarrow \nit{Course}(u,x, y), \nit{Dept}(y,x)$, where the last body atom appended to the original query is the {\em residue}
from that interaction, via resolution. This is reminiscent of {\em semantic query optimization} \cite{minker}, where satisfied ICs are used to optimize query answering, and also of {\em consistent query answering}
\cite[sec. 3.1]{2011Bertossi},
where possibly not satisfied ICs are imposed on queries to obtain semantically correct answers.
\boxtheorem
\end{example}
The procedure shown in the example could be refined to obtain contingent tuples for the obtained cause. Furthermore, it could be applied with Datalog extended with stratified negation \cite{Abiteboul95,ceri90}, using
{\em negation-as-failure} \cite{lloyd87} in the refutation. It could even be applied with causes for answers to conjunctive queries with negated atoms,\footnote{ \ This kind of queries were considered in
\cite{tapp16}, with a probabilistic approach.} and {\em Why-No} causes (as opposed to our {\em Why-So} causes \cite{Meliou2010a}), i.e. for not obtaining an expected  answer. This
could  be treated through view {\em insertions} with ICs, for which abduction can also be applied \cite{Console95}.

It is outside the scope of this work to give a full deductive-abductive approach to causes for answers to Datalog queries. However, it is worth mentioning that a FO, classical abductive approach
to view updates in the presence of ICs is proposed in \cite{Console95}. Continuing with our ongoing example, we briefly  sketch this approach.

\begin{example} (ex. \ref{ex:lp} cont.) \label{ex:fo} Consider again the query $\mc{Q}$ defined by (\ref{eq:heads}). Now, in FO-logic it becomes:\footnote{ \ Notice that this is the {\em completion}  of predicate $\nit{Ans}_{\mc{Q}}$ as defined by (\ref{eq:heads}). Predicate completion \cite{lloyd87} can be used  to deal
with more complex Datalog queries \cite{Console95}.}
\begin{equation}
\forall x(\nit{Ans}_{\mc{Q}_1}(x) \equiv \exists y \exists z(\nit{Dep}(y,x) \wedge
\nit{Course}(z,x,y))). \label{eq:compl}
\end{equation}
In contrapositive, considering that we want to virtually delete the unintended answer $\nit{Ans}_{\mc{Q}_1}({\sf John})$:
 \begin{eqnarray}
\neg \nit{Ans}_{\mc{Q}_1}({\sf John}) \ \equiv \ \forall y \forall z(\neg \nit{Dep}(y,{\sf John}) \vee
\neg \nit{Course}(z,{\sf John},y)). \label{eq:headsFOneg}
\end{eqnarray}
The formula on the right-hand side is (essentially) in disjunctive normal form (DNF), and expressed in terms of base atoms ( or abducible atoms).
It is obtained through the negation (due to a virtual answer deletion) of the (only partially ground)  {\em lineage} of the instantiated query \cite{probDBs,BunemanKT01,Karvounarakis12}.

Up to this point the ICs have not been taken into account. This is the next step. First, the IND is written in DNF as well, via Skolemization, obtaining
\begin{equation}
\psi'\!: \ \ \forall x \forall y \ (\neg \nit{Dep}(x, y) \vee \nit{Course}(f(x,y), y, x)), \label{eq:ind'}
\end{equation}
which is equiconsistent with $\psi$ \cite{lloyd87}. Next, to enforce the ICs, the atoms in (\ref{eq:headsFOneg}) are appended {\em residues} from the ICs. They are obtained by resolution between each of the
atoms (or more generally, literals) in (\ref{eq:headsFOneg}) and the constraint (\ref{eq:ind'}). In this case, $\nit{Dep}(y,{\sf John})$ has not residue, but $\nit{Course}(z,{\sf John},y))$
has $\nit{Dept}(y,{\sf John})$.\footnote{ \ Notice the similarity with query rewriting for obtaining {\em consistent query answers} from possibly inconsistent databases \cite[sec. 3.1]{2011Bertossi}.} So, the RHS of (\ref{eq:headsFOneg}) becomes:
 \begin{equation}
\forall y \forall z(\neg \nit{Dep}(y,{\sf John}) \vee
(\neg \nit{Course}(z,{\sf John},y) \wedge \nit{Dept}(y,{\sf John})). \label{eq:headsFOneg'}
\end{equation}
We could call the right-hand side the  {\em semantic lineage} of the (negated) query. \ Actually, it holds: \ $\mbox{(\ref{eq:compl}) } \wedge \mbox{ (\ref{eq:headsFOneg'}) }  \models \neg \nit{Ans}_{\mc{Q}}({\sf John})$
\cite{Console95}.
Notice that (\ref{eq:headsFOneg'}) can be written as:
 \begin{equation}
\forall y (\neg \nit{Dep}(y,{\sf John}) \vee (\neg \exists z \nit{Course}(z,{\sf John},y) \wedge \nit{Dept}(y,{\sf John}))). \label{eq:headsFOneg''}
\end{equation}
Due to the IND, the second disjunct (which is its negation) can be eliminated, simply obtaining: \ $\forall y \neg  \nit{Dep}(y,{\sf John})$.

Up to now the (extensional) database $D$ has not been considered. \ By looking it up, we obtain that the (minimal) abductive explanation is $\nit{Dep}({\sf Computing},{\sf John})$, leading to its deletion,
and to it as a cause for the original answer.

In our case, formula (\ref{eq:headsFOneg''}) is very simple. In general, it can be much more complicated, e.g. when we have: (a) More complex  Datalog queries, possibly with stratified negation, for which the intentional
predicate completions
have to be computed. In particular, conjunctive queries with negated atoms. \ (b) Several, possibly interacting ICs. \ (c) Complex view (intentional) updates, with both positive and negative ground atoms \cite{Console95}. For tuple view insertions denial constraints, in particular
key constraints and FDs, become relevant. It is possible to apply resolution to them, to obtain residues for the lineage literals.

The final interaction with the extensional
database $D$, to keep everything in a classical FO setting, can be done (via resolution and the {\em unique names assumption} \cite{lloyd87}) with the {\em logical reconstruction} of $D$ \cite{reiter82}. In our example, it is given by the theory:
 \begin{eqnarray*}
\forall x \forall y(\nit{ Dep}(x,y)&\equiv&(x= {\sf Computing} \wedge y ={\sf John}) \ \vee\\   &&(x= {\sf Philosophy} \wedge y=  {\sf Patrick}) \ \vee\\
&&~\vee \ (x = {\sf Math}  \wedge y=  {\sf Kevin})).\\
\forall x \forall y \forall z(\nit{Course}(x,y,z)  &\equiv& ( x = {\sf Com08} \wedge  y = {\sf John}  \wedge z =  {\sf Computing}) \ \vee \\
&&(x = {\sf Math01} \wedge y = {\sf Kevin}  \wedge z = {\sf Math}) \ \vee \\
&&(x ={\sf Hist02}\wedge  y = {\sf Patrick}   \wedge z = {\sf Philosophy}) \ \vee \\
&&(x ={\sf Math08}\wedge  y ={\sf Eli}   \wedge z ={\sf Math} ) \ \vee \\
&&(x = {\sf Com01}\wedge y= {\sf John} \wedge z = {\sf Computing})). \hspace{0.7cm} \mbox{\boxtheorem}
\end{eqnarray*}
\end{example}

 \section{Discussion and Conclusions}\label{sec:disc}

 In this work we have investigated the computational aspects causality for answers to Datalog queries. This was made possible by establishing a precise connections (mutual reductions) with
 adbuction from Datalog theories. This connection is interesting {\em per se}. In particular, the notion of necessity-degree for  abductive explanations, motivated by causality concepts, has been identified as relevant (cf. Section \ref{sec:nessDeg}).

We have also investigated in detail the connections between query-answer causality for monotone queries and updates through views defined by monotone queries. Particularly relevant is our investigation
of view-conditioned causality, for which we established connections with the view side-effect free delete propagation problem. We obtained new complexity results for both problems.

The problem of causality under integrity constraints (ICs) had not been investigated so far. Here we proposed the corresponding notions and obtained first complexity results. Abduction under
ICs was shown to be a promising direction to compute causes under ICs. There are still many problems and issues to investigate around causality in the presence of ICs.

  In this work we concentrated on {\em Why-So} causes, i.e. causes for obtained query answers. In \cite{Meliou2010a}, causality for {\em non-query-answers}, i.e. causes for {\em not} obtaining an expected answer, i.e.
  {\em Why-No} causality, is defined on basis of sets of {\em potentially missing tuples} that account for the missing answer. However, concepts and techniques for abduction under ICs as found in \cite{Console95} and suggested
  in Section \ref{sec:abdIC} seem to be applicable to {\em Why-No} causality. This is also left for future work.

In the rest of this section we discuss in a bit more depth some issues that deserve being considered for future research. At the same time we also mention some  related work that could be explored in more depth, for possibly
interesting connections with our work.

\subsection{Causality and ICs}

Some ICs are implicative, e.g. INDs and tgds, which makes it tempting to give them a causal semantics. For example,
in \cite{Roy14} and more in the context of interventions for explanations, a ground instantiation, $P_i(\bar{t}_i) \rightarrow P_j(\bar{t}_j)$, of an inclusion dependency is regarded a causal dependency of
$P_j(\bar{t}_j)$ upon $P_i(\bar{t}_i)$. On this basis, a {\em valid intervention}  removes  $P_j(\bar{t}_j)$ whenever  $P_i(\bar{t}_i)$ is removed from the instance. \re{This is in line with our general approach, as can be seen
from Example \ref{ex:ICex2}, with query $\mc{Q}_1$ and tuple $t_1$.}

Giving to ICs a causal connotation is controversial. Actually, according to \cite{Halpern10} logical dependencies
are not causal dependencies {\em per se}. \re{Our approach is also consistent with this view, in that antecedents of implications are not actual causes, but only elements of contingency sets, as can be seen, again, from Example \ref{ex:ICex2}, with query $\mc{Q}_1$ and tuple $t_1$}.

Our use in Section \ref{sec:abdIC} of the {\em semantic lineage} for determining causes in the presence of ICs leads, after grounding, to Boolean formulas in DNF. This opens the ground for possible applications
of  {\em knowledge compilation techniques} that are used in knowledge representation \cite{darwiche}, and had also provided interesting results in data management \cite{suciu}. This is direction that
deserves investigation.

Even more, we should point out that there are different
ways of seeing ICs, and they could have an impact on the notion of cause. For example, according to \cite{Reiter92},  ICs are ``epistemic in nature",
in the sense that rather than being statements about the domain represented by a database (or knowledge base), they are
statement about the {\em contents} of the database, or about what it {\em knows}.

\subsection{Endogenous tuples and view updates}\label{sec:viewsEnd}

The partition of a database into endogenous and exogenous tuples
used in causality may also be of interest in the context of delete-propagation. It makes sense to consider solutions based on {\em endogenous delete-propagation},  obtained through deletions
of endogenous tuples only. Actually, given an instance \re{$D=D^n \cup D^x$}, a view $ \mc{V}$ defined by a monotone query $\mc{Q}$, and  $\bar{a} \in \mc{V}(D)$, endogenous delete-propagation solutions for  $\bar{a}$ (in all of its flavors)  can be obtained from actual causes for $\bar{a}$ from the partitioned instance.

\begin{example} \label{exa:dppar} \ \re{(ex. \ref{ex:dpp} cont.)} \ Assume again that  $\langle\nit{{\sf John},{\sf XML}}\rangle$ has to be deleted from the query answer (view extension). Assume now only the data in the \nit{Journal} relation are reliable. Then, only deletions from the \nit{Author} relation make sense.  This can be captured by making \nit{Journal}-tuples exogenous, and \nit{Author}-tuples
endogenous. With this partition, only $ \nit{Author({\sf John}, {\sf TODS})}$ and $\nit{ Author({\sf John},{\sf TKDE})}$ are
  actual causes for $\langle\nit{\sf John},{\sf XML}\rangle$, with contingency sets $\Gamma=\{\nit{Author({\sf John}, {\sf TKDE})}\} $  and
 $\Gamma'=\{\nit{Author({\sf John}, {\sf TODS})}\}$, respectively  (see Example \ref{ex:cfex1}).

Now, each actual cause for $\langle\nit{{\sf John},{\sf XML}}\rangle$, together with its one-tuple subset-minimal (and also minimum-cardinality) contingency set, leads to the same set \{\nit{Author({\sf John},{\sf TODS}), Author({\sf John},{\sf TKDE})}\}, which, according to Propositions  \ref{pro:causeSubSDP} and \ref{pro:causeminSDP}, is
an  endogenous minimal- (and minimum-) delete-propagation solution
for $\langle\nit{{\sf John},{\sf XML}}\rangle$.
\boxtheorem
\end{example}

\subsection{Related connections}

 Our work, in combination with the results reported in \cite{tocs15}, shows that there are deeper and multiple connections between the areas
of QA-causality, abductive and consistency-based diagnosis, view-updates, and database repairs. Connections between consistency-based and abductive diagnosis have been established, e.g. in \cite{ConsoleT91}.
Abduction has  been explicitly applied to database repairs \cite{arieli}. The idea, again, is to ``abduce" possible repair updates that
bring the database to a consistent state.
Further exploring and exploiting these  connections is matter of ongoing and future research.

The view-update problem has been treated from the point of view of abductive reasoning
\cite{Kakas90,Console95}. The basic idea is to ``abduce" the presence of tuples in the base tables that explain the presence of those tuples in the view
extension, of those  one would like to, e.g. get rid of (cf. Section \ref{sec:abdIC}).

Database repairs are  related to the view-update problem.
Actually, {\em answer set programs} (ASPs) \cite{asp} for database repairs \cite[chap. 4]{2011Bertossi}  implicity repair the database by updating conjunctive combinations of intentional,
annotated predicates. Those logical combinations -views after all- capture violations of integrity constraints in the original database or along the (implicitly iterative) repair process
(a reason for the use of annotations).

In order to protect sensitive information, in \cite{lechen} databases are explicitly and virtually ``repaired" through secrecy views that specify the
information that has to be kept secret. In order to protect
information, a user is allowed to interact only with the virtually repaired versions of the original database that result from making those views empty or
contain only null values. Repairs are specified and computed using ASP, and an explicit connection to prioritized attribute-based repairs \cite{2011Bertossi}.

\vspace{2mm}
\noindent {\bf Acknowledgments:} \ Research funded by NSERC Discovery, and
the NSERC Strategic Network on Business Intelligence (BIN). We appreciate the feedback from anonymous reviewers for this paper, and from those for \cite{uai15,flairs16}.



\appendix

\section{Proofs of Results}\label{ap:proofs}

\defproof{Proposition \ref{pro:nessp}}{Consider a { DAP} $\mc{AP}= \langle \Pi, E, \nit{Hyp}, \nit{Obs}\rangle$ associated to $\Pi$, and $h \in \nit{Hyp}$. From
 the subset minimality of abductive diagnosis and Definition \ref{def:DAP} (part (c)), we obtain
  $h \in \nit{Ness}(\mc{AP})$ iff $\nit{Sol}(\mc{AP}')=\emptyset$ where, $\mc{AP'}= \langle \Pi, E, \nit{Hyp}
  \smallsetminus \{h\}, \nit{Obs}\rangle$. To decide whether $\nit{Sol}(\mc{AP}')=\emptyset$, it is
  good enough to check if $\Pi \cup E \cup \nit{Hyp} \models \nit{Obs}$. This can be done in polynomial time since Datalog
  evaluation is in  polynomial time in data complexity.}

 \defproof{Proposition \ref{pro:relp}}{{\em Membership:} Consider a Datalog abduction problem $\mc{AP}$ and a hypotheses $h \in Hyp$. To check whether $h$ is relevant for $\mc{AP}$, non-deterministically guess a subset $\Delta \subseteq\nit{Hyp}$, check if: (a) $h \in \Delta$, and (b) $\Delta$ is
 an abductive diagnosis for $\mc{AP}$. If $h$ passes both tests then it is  relevant, otherwise, it is irrelevant.

 Clearly, test
 (a) can be performed in polynomial time. We only need to show that checking (b) is also polynomial time. More  precisely, we need to show that $ \Pi \cup E \cup \Delta \models \nit{Obs}$ and $\Delta$ is subset-minimal. Checking whether $ \Pi \cup E \cup \Delta \models \nit{Obs}$ can be done in polynomial time, because Datalog evaluation is polynomial time. It is easy to verify that to check the minimality of $\Delta$, it is good enough to show that for all elements $\delta \in \Delta$, $ \Pi \cup E \cup \Delta \smallsetminus \{\delta\} \not \models \nit{Obs}$. This is because positive Datalog is monotone.

{\em Hardness:} We show that the combined complexity of deciding relevance for the Propositional Horn Clause Abduction (PHCA) problem, that is {\em  {NP}}-complete \cite{Friedrich90}, is a lower bound for the data complexity of the relevance problem for Datalog abduction.


A PHCA problem is of the form $\mc{P} =\langle \nit{Var, \mc{H}, SD, \mc{O}}\rangle$, where $\nit{Var}$ is a finite set of propositional variables, $\mc{H} \subseteq \nit{Var}$ contains
hypotheses, $\nit{SD}$ is a set of definite propositional Horn clauses, and $\mc{O} \subseteq \nit{Var}$ is the observation,  with $\mc{H} \cap \mc{O} = \emptyset$. An abductive diagnosis  for $\mc{P}$ is a subset-minimal $\Delta \subseteq \mc{H}$, such that  $\Delta \cup \nit{SD} \models \bigwedge_{o \in \mc{O}}o$. Deciding whether $h \in \mc{H}$ is relevant to $\mc{P}$ (i.e. it is an element of an abductive diagnosis of $\mc{P}$) is {\em {NP}}-complete \cite{Friedrich90}.

Deciding relevance for PHCA remains {\em NP}-hard for the {\em 3-bounded case} where: $\nit{SD}$ contains a rule
``$\nit{true} \leftarrow$", and all the other rules are of the form ``$a \leftarrow b_1, b_2, b_3$".\footnote{ \ Every  PHCA can be transformed to an equivalent 3-bounded PHCA,
because each rule  $a \leftarrow b_1, b_2, \ldots, b_n$
can be equivalently replaced by two rules $a \leftarrow c, \ldots, b_n$ and $c \leftarrow b_1, b_2$. Furthermore, $\nit{true}$ can be used to augment rule bodies with less than three propositional variables.}

Now, we provide a polynomial-time reduction from the problem of deciding relevance for 3-bounded PHCA
to our problem RLDP. To obtain data complexity for the latter, we need a fixed relational schema and a fixed Datalog program $\Pi$ over it, so that inputs for relevance in 3-bounded PHCA are mapped to
the extensional components of $\Pi$, where relevance is tested. 

More precisely, given a 3-bounded PHCA $\mc{P}$, build the DAP problem $\mathcal{AP^\mc{P}}= \langle \Pi, E^\mc{P}, \nit{Hyp}^\mc{P},   \nit{Obs}^\mc{P}\rangle$ as follows, where
 $\Pi$ is the following (non-propositional) Datalog program (whose underlying domain consists of the propositional variables in $\nit{SD}$ plus $\nit{true}$):
\begin{eqnarray}
 T(\nit{true})&\leftarrow& \label{eq:red1}\\
 T(x_0)&\leftarrow& T(x_1), T(x_2), T(x_3), R(x_0, x_1, x_2, x_3).\label{eq:red2}
\end{eqnarray}
Furthermore, $E^\mc{P} := \{R(a, b_1, b_2, b_3)~|~ a \leftarrow b_1, b_2, b_3 \mbox{ appears in } \nit{SD}\}$. Furthermore, $\nit{Hyp}=\{T(a) \  | \ a \in \mc{H} \}$ and $\nit{Obs}=\{T(a) \  | \ a \in \mc{O} \}$.
Notice that this reduction can be done in polynomial-time in the size of $\mc{P}$.

It is possible to prove that: \ For a $\mc{P} =\langle \nit{Var, \mc{H}, SD, \mc{O}}\rangle$ and a hypothesis $h \in \mc{H}$,
$h$ is relevant for $\mc{P}$ iff $T(h) \in \nit{Rel}(\mathcal{AP^\mc{P}})$.}

\ignore{\begin{lemma} \em \label{lem:reduc} For a $\mc{P} =\langle \nit{Var, \mc{H}, SD, \mc{O}}\rangle$ and a hypothesis $h \in \mc{H}$,
$h$ is relevant for $\mc{P}$ iff $T(h) \in \nit{Rel}(\mathcal{AP^\mc{P}})$.
\boxtheorem
\end{lemma} }

The following example illustrates the reduction in the hardness part of the proof of Proposition \ref{pro:relp}.

\begin{example} Consider the ``Propositional Horn Clause Abduction" (PHCA) $\mc{P} =\langle   \{a,b,c\}, \{c,b\}, \{ a \leftarrow b,c \ ; \ b  \leftarrow c  \}, \{a\}\rangle$,
whose components are, respectively, a set of propositional variables, a subset of the former formed by the abductibles (hypothesis), a positive propositional program, and the set of
observations.
It is easy to verify that $\mc{P}$ has the single abductive diagnosis, $\{c\}$, and then  a single relevant hypotheses, $c$.

The 3-bounded PHCA $\mc{P}^{3b} =\langle   \{a,b,c\}, \{c,b\},  \{ \nit{true}\leftarrow \ ; \ a \leftarrow b,c, \nit{true} \ ; \  b  \leftarrow c, \nit{true}, \nit{true}  \}, \{a\}\rangle$ is equivalent to $\mc{P}$.

Now, $\mc{P}^{3b}$ can be mapped to the DAP  $\mathcal{AP}^{\mc{P}^{3b}}= \langle \Pi, \{ R(a,b,c,  \nit{true}),$ \linebreak $R(c,b,\nit{true}, \nit{true})\}, \{ T(c), T(b)\}, \{ T(a))\} \rangle $, with $\Pi$ as
in (\ref{eq:red1}, (\ref{eq:red2}), which
has a single abductive diagnosis, $\{T(c)\}$. 
\boxtheorem
\end{example}


\ignore{
\defproof{ Proposition \ref{pro:relp}}{ To show the membership to {\em  {NP}}: given a Datalog abduction $\mc{AP}$ and $h \in Hyp$, non-deterministically guess a subset $\Delta \subseteq\nit{Hyp}$, check if a) $h \in \Delta$ and b) $\Delta$ is an abductive diagnosis for $\mc{AP}$, then $h$ is relevant, Otherwise, it is not relevant. Obviously, (a) can be checked in polynomial. We only need to show that checking (b) is also polynomial. More  precisely, we need to show that $ \Pi \cup E \cup \Delta \models \nit{Obs}$ and $\Delta$ is subset-minimal. Checking whether $ \Pi \cup E \cup \Delta \models \nit{Obs}$ can be done in polynomial, because Datalog evaluation is polynomial time. It is easy to verify that to check the minimality of $\Delta$, it is good enough to show that for all elements $\delta \in \Delta$, $ \Pi \cup E \cup \Delta \not \models \nit{Obs}$. This is because positive Datalog is monotone. Therefore, relevance problem belongs to {\em  {NP}}.

We establish the hardness by showing that the combined complexity of the relevance problem for Propositional Horn Clause Abduction ( {PHCA}), shown to be {\em  {NP}}-complete in \cite{Friedrich90}, is an lower bound for the data complexity of the relevance problem for Datalog abduction.  {PHCA} is a tuple $P =(\nit{Var, \mc{H}, SD, \mc{O}})$, where $\nit{Var}$ is a finite set of propositional variables, $\mc{H} \subseteq \nit{Var}$ are
the individual Hypotheses, $\nit{SD}$ is a set of definite propositional Horn clauses, and $\mc{O} \subseteq \nit{Var}$  the observation, is a finite conjunction of propositions with $\mc{H} \cap \mc{O} = \emptyset$. An abductive diagnosis  for $P =(\nit{Var, \mc{H}, SD, \mc{O}})$ is a subset-minimal $\Delta \subseteq \mc{H}$ such that  $\Delta \cup \nit{SD} \models Obs$. It is known that deciding whether $h \in \mc{H}$ is relevant to $P$ (i.e., it is an element of an abductive diagnosis of $P$) is {\em {NP}}-complete.

We call a  {PHCA} 3-bounded if its rules are of the form  $\nit{true} \leftarrow$  or $a \leftarrow b_1, b_2, b_3$. It is clear that all  {PHCA}s can be converted to an equivalent 3-bounded  {PHCA}. Without loss of generality, assume $|\mc{O}|=1$.

For a 3-bounded  {PHCA}, we define a Datalog abduction  $\mathcal{AP}= \langle \Pi, E, \nit{Hyp}, \nit{Obs}\rangle$ where, $\Pi$ is
\begin{eqnarray*}
t(true)&\leftarrow&\\
t(x_0)&\leftarrow& t(x_1), t(x_2), t(x_3), r_i(x_0, x_1, x_2, x_3), \\
\end{eqnarray*}

\vspace{-.7cm}
and the input structure contains a ground atom $r_i(a, b_1, b_2, b_3)$ iff the rule $a \leftarrow b_1, b_2, b_3$ occurs in $\nit{SD}$. Furthermore, $\nit{Hyp}=\{ t(x)| \ x \in \mc{H} \}$ and $\nit{Obs}=\{t(x)| \ x \in \mc{O} \}$.

It is not difficult to verify that $h$ is a relevant hypothesis for a  {PHCA} $P$ iff $t(h)$ is relevant for the corresponding Datalog abduction $\mc{AP}$.   $\mc{AP}$ has a fixed program and the size of $\nit{SD}$ is pushed to the size of the input structure $\nit{EDB}$ i.e., combined complexity of the relevance problem for  {PHCA} is a lower bound for the data complexity of Datalog abduction. Therefore, relevance problem is {\em  {NP}}-hard.}
}

\ignore{
\defproof{Proposition \ref{pro:abdf&cfcaus}}{ Fist, assume $\tau$ is an actual for \nit{ans}. According to Definition \ref{def:querycause} (slightly modified for Datalog queries) there exists a contingency set  $ \Gamma \subseteq D^n$ s.t. $ \Pi \cup D \smallsetminus  \Gamma \models  \nit{ans}$ but $\Pi \cup D \smallsetminus \Gamma - \{t\} \not \models  \nit{ans}$. This implies that there exists a set $\Delta \subseteq D^n$ with $t \in \Delta $ s.t.   $ \Pi \cup \Delta \models \nit{ans}$. It is easy to see that  $ \Delta$ is an abductive diagnosis for $\mathcal{AP}^c$. Therefore, $t \in \nit{Rel}(\mathcal{AP}^c)$.

Second, assume $t \in \nit{Rel}(\mathcal{AP}^c)$. Then there exists a set $\mc{S}_k \in \nit{Sol}(\mathcal{AP}^c)=\{s_1 \ldots s_n\}$ such that $\mc{S}_k \models \nit{ans}$ with $t \in \mc{S}_k$. Obviously, $\nit{Sol}(\mathcal{AP}^c)$ is a collection of subsets of $D^n$. Pick a  set $\Gamma \subseteq D^n$ s.t., for all $\mc{S}_i \in \nit{Sol}(\mathcal{AP}^c)$ $i \not = k$, $\Gamma \cap \mc{S}_i \not = \emptyset$ and  $\Gamma \cap \mc{S}_k =\emptyset$. It is clear that $ \Pi \cup D \smallsetminus \Gamma -\{t\} \not \models \nit{ans}$ but  $\Pi \cup D \smallsetminus \Gamma \models \nit{ans}$. Therefore, $\tau$ is an actual cause for \nit{ans}.
To complete the proof we need to show that such $\Gamma$ always exists. This can be done by applying the digitalization technique to construct such $\Gamma$. Since all elements of $\nit{Sol}(\mathcal{AP}^c)$ are subset-minimal, then, for each $\mc{S}_i \in \nit{Sol}( \mathcal{AP}^c)$ with $i \not = k$, there exists a $t' \in \mc{S}_i$ such that $t' \not \in \mc{S}_k$.  Therefore, $\Gamma$ can be obtained from the union of difference between each $\mc{S}_i$ ($i \not = k$) and $\mc{S}_k$.
}
}

\ignore{
\defproof{Proposition \ref{pro:abdf&res}}{
Assume $N$ is a minimal cardinality set of necessary
hypothesis set of $\tau$. From the definition a necassry hypothesis set, it is clear that $\Gamma= N-\{t\}$ is a cardinality minimal contingency set for $\tau$ and the result is followed.
}
}

\ignore{
\defproof{Proposition \ref{pro:causeSubSDP}}{ The results is simply follows from the definition of an actual cause.  Assume a set $D' \subset D$ and $t \in  D \smallsetminus D' $.  If $D'$ is a solution to a source minimal side-effect deletion-problem then $\tau$ is an actual cause for $a$ with contingency  $D \smallsetminus D' -\{t\}$. Likewise, removing each actual cause for $a$ together with one of it S-minimal contingency set is a solution to source minimal
side-effect deletion-problem.
}
}

\ignore{
\defproof{Proposition \ref{pro:cp}}{  To show the membership to \nit{ {NP}}:  non-deterministically guess a subset $\Gamma \subseteq D^n$, return yes if $D\cup  \Gamma \cup \{t\} \not \models \mc{Q}(\bar{a})$ and $\Gamma \cup \{t\} \models \mc{Q}(\bar{a}) $.  Otherwise no. Checking the mentioned conditions consists of  two Datalog query evaluations that is polynomial in data. The \nit{ {NP}}-hardness is obtained by the reduction from the relevance problem for Datalog abduction to causality problem provided in \ref{pro:ac&rel}.}
}

\ignore{
\defproof{Proposition \ref{pro:causefromviewII}}{ The \nit{ {NP}}-hardness is obtained by the reduction from the view-side-effect free problem for Datalog abduction to  {vc-}cause problem provided in \ref{pro:vc&view}. Membership to \nit{ {NP}} can be shown similar to that of \ref{pro:cp} }
}

\end{document}